\newcounter{myctr}
\def\myitem{\refstepcounter{myctr}\bibfont\noindent\ifnum\themyctr>9\else\phantom{0}\fi\hangindent17pt\themyctr.\enskip}

\documentclass{ws-ijqi}
\DeclareGraphicsExtensions{.pdf}

\begin{document}

\markboth{Fitzpatrick, Simon, Sergienko} 
{High-Cap. Img. Rot. Insen. Obj. Id. with Correlated OAM States}


\title{High-Capacity Imaging and Rotationally Insensitive Object Identification with Correlated Orbital Angular Momentum States} 

\author{CASEY A. FITZPATRICK\footnote{Corresponding author.}}

\address{Department of Electrical and Computer Engineering, Boston University, 8 St. Mary's St.,\\
Boston, MA 02215, United States of America\\
cfitz@bu.edu}
\author{DAVID S. SIMON}
\address{Department of Electrical and Computer Engineering, Boston University, 8 St. Mary's St.,\\
Boston, MA 02215, United States of America\\  Dept. of Physics and Astronomy, Stonehill College, 320 Washington Street, \\ Easton, MA 02357, United States of America \\
simond@bu.edu}

\author{ALEXANDER V. SERGIENKO}

\address{Department of Electrical and Computer Engineering, Photonics Center, Boston University, 8 St. Mary's St.,
Boston, MA 02215, United States of America\\  Dept. of Physics, Boston University, 590 Commonwealth Ave., \\ Boston, MA 02215 \\
alexserg@bu.edu}

\maketitle

\begin{history}
\received{15 September 2014}
\revised{Day Month Year}
\end{history}

\begin{abstract}
Using no conventional measurements in position space, information extraction rates exceeding one bit per photon are achieved by employing high-dimensional correlated orbital angular momentum (OAM) states for object recognition. The correlations are shown to be insensitive to axial rotation of the target object: the information structure of an object's joint OAM coincidence spectrum is unchanged even when the object undergoes random rotations between each measurement. Additionally, OAM correlations alone are shown to be sufficient for full image reconstruction of complex, off-axis objects, and novel object symmetries are observed in the phases of OAM-object interaction transition amplitudes. Variations in mutual information rates, due to off-axis translation in the  beam field, are studied, and it is shown that object symmetry signatures and information rates are independent of environmental factors sufficiently far from the beam center. The results motivate dynamic scanning applications in contexts where symmetry and small numbers of noninvasive measurements are desired. 
\end{abstract}


\section{Introduction}\label{introsection}	

Recently, a new method of correlated optical sensing was introduced \cite{1}, in which correlations between the orbital angular momentum (OAM) states \cite{2,3,4} of two beams or photons are measured through coincidence counting and singles rates. If an object is placed in one of the paths, the correlations are effected in ways that are useful for efficient sensing applications. As we shall see, the sensing procedure employs the inherent rotational symmetries of OAM modes in order to increase the information per photon measured, while decreasing the number of measurements needed to identify or even {\it image} target objects. All correlations used are based on count rates, so that no measurements in position space are needed. (See \cite{5,6,7,8,9,10,11,12,13,14,15,16,17} for related, {\it non} OAM-based methods such as ghost imaging and compressive ghost imaging, which require spatially-resolving measurements.) By relying on the OAM correlations explored below, simple robust measurements may be used to extract more information for less energy as compared to traditional photon-based sensing applications.

Aside from predicting image reconstruction capabilities, the method explored in \cite{1} suggested that objects leave an imprint in the off-diagonal components of the joint OAM coincidence spectrum, where the diagonal represents the well-known conservation of OAM \cite{18}. The latter was confirmed experimentally for simple cases in \cite{19}. In these experiments it was observed that, for centered objects, the joint OAM spectrum signifies and object's basic rotational symmetries in a predictable way, namely that only the  off-diagonal components of the spectrum for which $l_o + l_r = N$ are nonzero when the object transmission pro has an $N$-fold rotational symmetry, where $l_o$ and $l_r$ represent the OAM content of the object and reference beam, respectively. The theory introduced in \cite{1} also suggests a potential imaging scheme based on correlated OAM states, which we will hereafter refer to as {\it correlated spiral imaging} (CSI) in reference to what is traditionally called {\it digital spiral imaging} \cite{22,23,24}, an interesting related but inherently non-imaging spectroscopic method. By contrast, CSI allows full image reconstruction by way of singles rates measurements, as well as object recognition by way of joint OAM coincidence spectrum measurement.

Here, we present CSI's novel ability to clearly identify arbitrary rotating objects, our numerical results illustrating the {\it rotational insensitivity} of this correlation method. In other words, we show that the coincidence spectra generated using our method remain unchanged even when the target object undergoes random rotations in between {\it each} measurement. Further, we generalize the domain of application of CSI by employing substantially more complicated objects than have been previously considered. By paying particular attention to image reconstruction as well as the mutual information dynamics of the coincidence spectrum associated to each object for object recognition, we characterize the most relevant parameters of CSI with greater completeness than has previously been considered.

By "more complicated" we mean objects with complex geometries, a great deal of angular variation, and at off-center positions in the beam field. These novel cases are mathematically non-trivial and, for reasons discussed below, currently experimentally infeasible. We are thus forced at present to seek further insight into the correlations discussed below by way of simulation. We use digitized representations of opaque objects to directly compute the complex transition amplitudes needed in order to study the effects -- on spectral signature, CSI reconstruction accuracy, and mutual information rates -- of linearly translating objects off-axis with respect to the beam's center. This process follows a proposed measurement scheme outlined below. It is important to reiterate that in all of what follows, no traditional imaging measurements in position space are assumed, only singles and coincidence rates, as they are required to properly express the complex expansion coefficients (transition amplitudes) used to reconstruct object images.This makes CSI particularly well-suited for applications where spatially resolving detectors are unavailable or undesirable.

After reviewing the relevant theoretical concepts in Sec.~\ref{theory} and discussing their relation to the CSI measurement apparatus in Sec.~\ref{apparatus}, we present a number of numerical results in Sec.~\ref{simmysec}. We show not only that the method successfully images asymmetric off-axis objects, but that the mutual information contained in the off-diagonal elements of the joint OAM spectrum remains above the one bit per photon limit for objects near the beam center, and grows with the size of the object's symmetry group. The study of the off-axis behavior is a prerequisite for use in any scanning-based applications. Additionally, we show for the first time (going beyond the predictions of \cite{1} and experiments of \cite{19}) that object symmetry properties are in fact present in the {\it phases} of the OAM-object interaction transition amplitudes (the image expansion coefficients). This unexpected result further highlights the importance of these heretofore unconsidered quantities. Finally, we show the rotational insensitivity of the coincidence spectra before concluding in Sec.~\ref{concludesection}

\section{Background}\label{background}
\subsection{Theory}\label{theory}

All beams considered will be decomposed into Laguerre-Gauss (LG) modes $|l,p\rangle$, each mode with OAM $l\hbar$ and $p$ radial nodes where $l$ and $p$ are integers \cite{20,21}. In cylindrical coordinates $(r,z,\phi)$, the position space representation of $|l,p\rangle$ in the object plane is given by \cite{26} \begin{align}\langle r,\phi | l,p\rangle = k^{|l|}_{p}r^{|l|}e^{-r^{2}/w_{0}^{2}}L^{|l|}_{p}(2r^{2}/w_{0}^{2})e^{-i\phi l + i(2p + |l| + 1)},  \end{align} where $k^{|l|}_{p}$ is a normalization constant, $L_n^{\alpha}(x)$ is the generalized Laguerre polynomial of order $n$, and $w_{0}$ is the beam waist \cite{4}. 

Before discussing the measurement apparatus, it will be useful to develop a sense of how objects may be represented in terms of their effect on OAM states. By considering this representation now, the role of correlation will become more clear later. We use outer products of OAM states to form an overcomplete basis which can be used to express an object's transmission function, $T(r,\phi)$, as a local operator expanded in terms of OAM transitions as follows. For some $a^{l^{'}l}_{p^{'}p} \in \mathbb{C}$  \begin{align} \hat T =  \sum_{l,l^{'},p,p^{'}} a^{l^{'}l}_{p^{'}p} |l^{'},p^{'}\rangle \langle l,p| .\label{lgobjeqn}\end{align} To better understand the coefficients $a^{l^{'}l}_{p^{'}p} $, we invert Eq.~(\ref{lgobjeqn}) and find that \begin{eqnarray} a^{l^{'}l}_{p^{'}p} &=& \int drd\phi [u_{l^{'}p^{'}}(r,\phi)]^{*}T(r,\phi)u_{lp}(r,\phi)\notag\\&=&\langle l^{'},p^{'} |\hat T |l,p \rangle \label{transamp}.\end{eqnarray} Eq. (\ref{transamp}) shows that the $a^{l^{'}l}_{p^{'}p}$ are in fact {\it amplitudes} for a given state $|l,p\rangle$ to transition to the state $|l^{'},p^{'}\rangle$. Since the LG state vectors are {\it a priori} known basis elements, successful reconstruction of an image depends only on the determination of the $a^{l^{'}l}_{p^{'}p}$, as indicated by Eq.~(\ref{lgobjeqn}). We note here a clear analogy with Fourier decomposition, with the exception that our basis functions form an overcomplete basis.

The $a^{l^{'}l}_{p^{'}p}$ are complex expansion coefficients according to this representation of the object. To properly represent the object then, both the magnitude \emph{and} phase (or, equivalently, the real and imaginary parts) of the $a^{l^{'}l}_{p^{'}p}$ are needed. Throughout this paper \emph{the phases we refer to are the relative phases between quantum amplitudes} of the sum in Eq.~\ref{lgobjeqn}, as opposed to for example, the phase profile of the object or spatial phase distribution of the OAM states themselves. It is crucial to observe, as can be seen from Eq.~\ref{lgobjeqn} alone, that the complete $a^{l^{'}l}_{p^{'}p} = \mathbb{\Re}(a^{l^{'}l}_{p^{'}p}) + i\mathbb{\Im}(a^{l^{'}l}_{p^{'}p}) = |a^{l^{'}l}_{p^{'}p}|e^{i\alpha}$ for some phase $\alpha$ describes the rotation (in complex space) of the basis state $|l^{'},p^{'}\rangle \langle l,p|$ necessary to give the proper interferences needed to accurately represent the object. It was pointed out in \cite{1} that without these full complex numbers all images become rotationally symmetric, thus destroying the potential for generalized imaging.

The \emph{magnitudes} of the $a^{l^{'}l}_{p^{'}p}$ in Eqs.~(\ref{lgobjeqn},\ref{transamp}) are proportional to the coincidence probabilities with a proportionality constant corresponding to the weighting coefficient from SPDC (\cite{1}, \cite{19}). From the preceding paragraph we see that, given these magnitudes, proper imaging of an object still requires the phases ($\alpha$'s) too, or just as well, the real and imaginary parts of the $a^{l^{'}l}_{p^{'}p}$ defined in Eq.~\ref{transamp}. The next section describes a method for measuring the real an imaginary parts of the $a^{l^{'}l}_{p^{'}p}$.

\subsection{Apparatus}\label{apparatus}

\begin{figure}
\centering
\includegraphics[totalheight=1.6in]{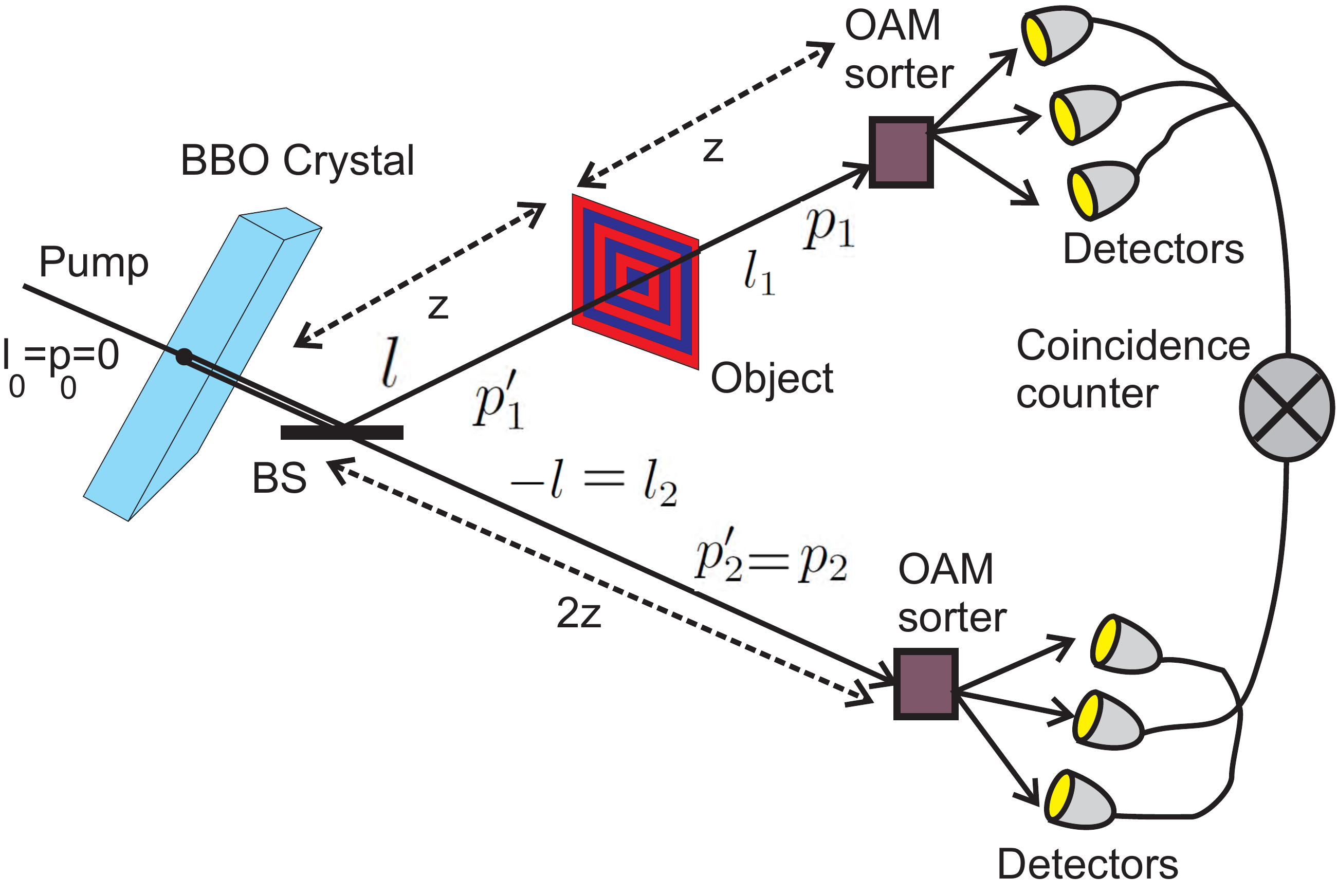}
\caption{ A phase-insensitive setup allowing measurement of correlated OAM content, but no image reconstruction.}
\label{setup1}\end{figure}

\begin{figure}
\centering
\includegraphics[totalheight=1.6in]{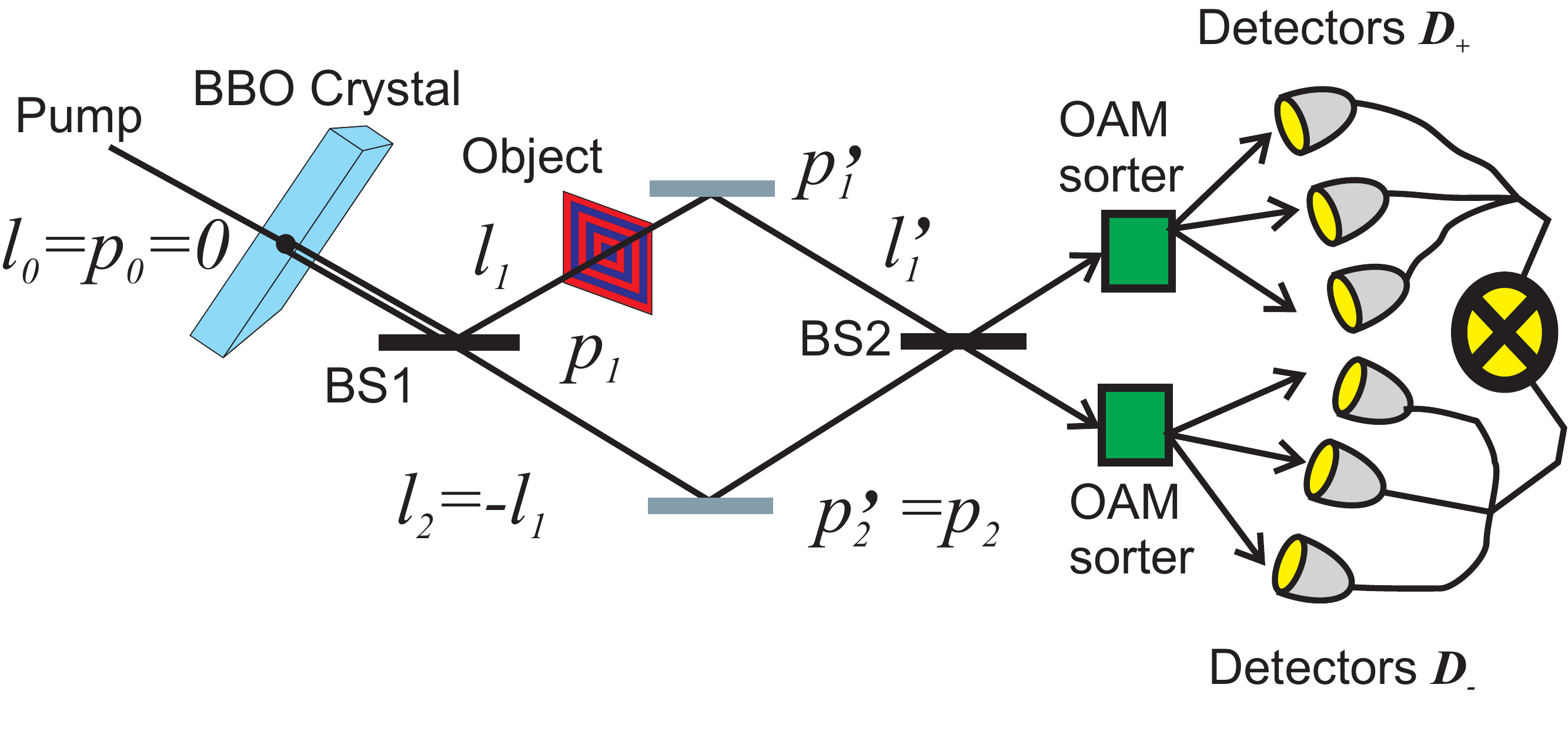}
\caption{ A setup allowing object image reconstruction via transition amplitude phase-sensitive measurement of correlated OAM content.}
\label{setup}\end{figure}

The phase-{\it insensitive} apparatus, originally described in \cite{1} and implemented in \cite{19} is shown in Fig.~\ref{apparatus}. From this setup, an object's joint OAM spectrum may be constructed, as demonstrated experimentally in \cite{19}. However, no true images can be built from the data output by this apparatus, because only the magnitudes, not the phases, of the OAM-object interaction amplitudes, $a^{l^{'}l}_{p^{'}p}$, are recoverable from the measurements. Below we show that the phases can be recovered by implementing a simple augmentation to the setup in Fig.~\ref{setup1}, shown in Fig.~\ref{setup}. Namely, the introduction of an additional beam splitter, along with the counting of singles rates in each detector, is all that is needed. The second beam splitter (BS2) introduces a path-ambiguity that can be mathematically exploited for full image reconstruction.

We will now explain how the transition amplitudes, $a^{l^{'}l}_{p^{'}p}$ (both real and imaginary parts) of Eq.~(\ref{transamp}), may be recovered using the correlation setup shown in Fig.~\ref{setup}. Recall that, according to Eq.~(\ref{lgobjeqn}), both the real and imaginary parts, or equivalently, the magnitude and phase of the  $a^{l^{'}l}_{p^{'}p}$ are necessary to provide the proper representation of the object. For further disambiguation of the term \emph{phase} in this context, see Sec.~\ref{theory}. 

As a source of photon pairs which are correlated in OAM, consider a Gaussian pump, $|l_{p},p_{0}\rangle=|0,p_{0}\rangle$, producing object and reference beams, $|l_{1},p_{1}\rangle$ and $|l_{2},p_{2}\rangle$ respectively, via (collinear) spontaneous parametric down-conversion (SPDC); type I or type II SPDC will suffice, however coincidence coincidence rates are higher for type II SPDC with a polarizing beam splitter \cite{24,25,26,27,28,1}. The beams, anti-correlated in OAM since $l_{p}=l_{1}+l_{2}$, are then directed into separate branches by a 50:50 beam-splitter (see Fig.~\ref{setup}). The presence of the object will cause the state of a photon in the object beam to transition from $|l_{1},p_{1}\rangle$ to $|l^{'}_{1},p^{'}_{1}\rangle$, while the state of the reference photon remains unchanged. According to Eqs.~(\ref{transamp}) and (\ref{lgobjeqn}), the object's \emph{OAM transition amplitude} is precisely what we seek to recover, and we do so as follows. 

By inserting an additional beam splitter before sorting the states, amplitude path-mixing is induced (see Fig.~\ref{setup}) because photons reaching either detector may have arrived by two different routes. Note that both ports of the second beam splitter are being used, so that vacuum fluctuations need not be taken into account here. The resulting detection {\it amplitudes} in the upper ($D_{+}$) and lower ($D_{-}$) detectors are, respectively, proportional to $|a^{l_{1}^{'}l_{1}}_{p_{1}^{'}p_{1}}+i|$ and $|ia^{l_{1}^{'}l_{1}}_{p_{1}^{'}p_{1}}+1|$, neglecting overall constants (Note that the amplitudes associated to $|l_{2},p_{2}\rangle$ are $1$ since no transition takes place, hence those quantum numbers do not appear in the beam splitter expressions.) The proportionality constant is determined by the weighting coefficients associated to the SPDC process \cite{26}. This means that the singles count rates $N_{+}$ and $N_{-}$ in each detector are given by \begin{eqnarray} N_{+}&\approx&|a^{l_{1}^{'}l_{1}}_{p_{1}^{'}p_{1}}+i|^{2} \\ N_{-}&\approx&|ia^{l_{1}^{'}l_{1}}_{p_{1}^{'}p_{1}}+1|^{2}.\end{eqnarray} We assume perfect detectors; imperfections  can be accounted for by methods described in \cite{26}. Considering the real and imaginary parts of the $a^{l^{'}l}_{p^{'}p}$ as the two unknown variables we seek, it is clear that we have two equations and two unknowns. Thus, the singles rate equations can now be used to express both the real and imaginary parts of the $a^{l_{1}^{'}l_{1}}_{p_{1}^{'}p_{1}}$. So, using the real part, $\mathbb{\Re}(a^{l_{1}^{'}l_{1}}_{p_{1}^{'}p_{1}})=\sqrt{N_{+}-1-\frac{1}{16}(N_{+}^{2}-N_{-}^{2}-2N_{+}N_{-})-\frac{1}{2}(N_{+}-N_{-})}$, and the imaginary part, $\Im(a^{l_{1}^{'}l_{1}}_{p_{1}^{'}p_{1}})=\frac{1}{4}(N_{+}-N_{-})$ of $a^{l_{1}^{'}l_{1}}_{p_{1}^{'}p_{1}}$,  we now have the $a^{l^{'}l}_{p^{'}p}$ in their entirety, and all necessary information for image reconstruction. Further, we have as a subset of this data the coincidences required to build the OAM joint spectrum used for object identification.

We would like to point out that while the amplitudes used for image reconstruction \emph{can} be used for construction of the OAM joint spectrum, we are not using the \emph{spectrum} to image; we are using the amplitudes of Eq.~\ref{lgobjeqn} to image. The spectrum measurement (without the second beam splitter) provides only the \emph{magnitudes} of the $a^{l^{'}l}_{p^{'}p}$, and not the \emph{phases} of the $a^{l^{'}l}_{p^{'}p}$ (as noted in Sec.~\ref{theory} we are \emph{not} referring to the phase profile of the object or OAM beam). However, as argued in Sec.~\ref{theory} and \cite{1} this phase information is needed for image reconstruction, and to obtain these phases we need both the real and imaginary parts of the $a^{l^{'}l}_{p^{'}p}$. The above apparatus provides a way of obtaining this information and thus imaging an object without making any measurements in position space. Additionally, the apparatus allows construction of the more familiar joint OAM spectrum of the object, studied for simple cases in \cite{19} and more general cases below.

Finally, let $P(l_1,p_1;l_2,p_2)$ be the mutual probability for detecting signal with quantum numbers $l_1,p_1$ and idler with values $l_2,p_2$. The marginal probabilities at the two detectors (probabilities for detection of a single photon, rather than for coincidence detection) are \begin{eqnarray}P_s(l_1,p_1)&=&\sum_{l_2,p_2} P(l_1,p_1;l_2,p_2)\\ P_i(l_2,p_2)&=&\sum_{l_1,p_1} P(l_1,p_1;l_2,p_2).\end{eqnarray} Then the mutual information for the pair is \begin{eqnarray}I(s,i)&=& \sum_{l_1,l_2=l_{min}}^{l_{max}} \sum_{p_1,p_2=0}^{p_{max}}P(l_1,p_1;l_2,p_2) \label{mutinfo}\\ & & \times \; \log_2 \left( {{P(l_1,p_1;l_2,p_2)}\over {P_s(l_1,p_1)P_i(l_2,p_2)}}\right)\nonumber \end{eqnarray}

The setup in Fig.~\ref{setup} can be used, accounting for the SPDC weighting coefficients, to generate OAM coincidence spectra, as we shall see in Sec.~\ref{simmysec}. Alternatively, one can simply remove the second beam splitter and measure the coincidences directly; images will not be attainable since only the transition amplitude magnitudes will be known, but the spectral signature of objects may still be obtained \cite{1,19}. Which version of the setup one uses will depend on the application at hand, in particular whether or not an image is desired or a spectral signature will suffice.

So far we have assumed the experimenter has the ability to discriminate different $p$ values while sorting OAM states. Such discrimination is \emph{necessary} in order to properly represent an object in terms of Eq.~\ref{lgobjeqn}. However, in reality such a discrimination is beyond current experimental means \cite{28}. Therefore, to demonstrate the full implications of the above detection method, we must (presently) rely on simulation methods, described in Sec.~\ref{simmysec}.

\section{Imaging and Information}\label{simmysec}

The following simulations explore correlated spiral imaging (CSI), a remote OAM-based imaging and object recognition technique that requires few measurements since OAM symmetries lead to information extraction rates exceeding one bit per photon measured. In addition to testing the imaging capabilities of CSI, we will see that an aspect of the technique, namely, the object joint OAM signature, is {\it insensitive to rotations of the object}, so that the mutual information advantages remain constant even for dynamic objects. This is perhaps the most advantageous and unusual property of CSI, opening the door for novel applications across many fields concerned with the identification and imaging moving objects.

\subsection{Complex Objects: Imaging and Joint OAM Spectra}\label{complexobjects}

Successful imaging of any object that has significant radial structure with CSI requires the ability to discriminate OAM states with $p>0$, since the expansion basis for $\hat T$ depends upon distinct contributions from each $(p^\prime, p)$ combination in the set of basis states (see Eq.~(\ref{lgobjeqn})). In the simulations below, following the physics of Fig.~\ref{setup}, we use digitized representations of various opaque objects to directly compute Eq.~(\ref{transamp}) in order to study the effects -- on spectral signature, image reconstruction accuracy, and mutual information -- of translating the target objects off-axis with respect to the beam's center. In addition to image reconstructions we show many joint OAM coincidence spectra and use these spectra to perform mutual information calculations using Eq.~(\ref{mutinfo}). 

As discussed Sec.~\ref{apparatus}, the spectra may be constructed from singles measurements combined with coincidence measurements and the addition of a second beam splitter before detection (as we simulate here), or, directly measured in a coincidence-only setup, with no second beam splitter, as in \cite{19}. These results may be tested experimentally as distinguishing states with $p>0$ becomes more practical.

The experimental results of \cite{19} briefly discuss the role of symmetry in an object's joint OAM spectrum. The most important point for our purposes is that objects with simple $N$-fold rotational symmetry leave their imprint on the {\it off-diagonal} (in particular, the $l_{o}+l_{r}=N$) elements of the joint spectrum. More generally, suppose that the object has a rotational symmetry group of order $N$; i.e., it is invariant under $\phi\to \phi +{{2\pi}\over N}$. From Eqs. (\ref{lgobjeqn}) and (\ref{transamp}) it follows that the coefficients must then satisfy $a_{p_1^\prime p_1}^{l_1^\prime l_1}=e^{{{2\pi i}\over N}(l_1^\prime -l_1)}a_{p_1^\prime p_1}^{l_1^\prime l_1}$, which implies $a_{p_1^\prime p_1}^{l_1^\prime l_1}=0$ except when ${{l_1^\prime -l_1}\over N}$ is integer. When $N$ goes up (enlarged symmetry group), the number of nonzero $a_{p_1^\prime p_1}^{l_1^\prime l_1}$ goes down; with the probability concentrated in a smaller number of configurations, correlations increase and mutual information goes up, as we will see below.

It is also worth noting that the objects used in \cite{19}, while having width much smaller than the beam waist, had length that extended far beyond the beam radius.

 \begin{figure}
 	\begin{minipage}[c]{0.48\columnwidth}
		 \includegraphics[height=1.5in]{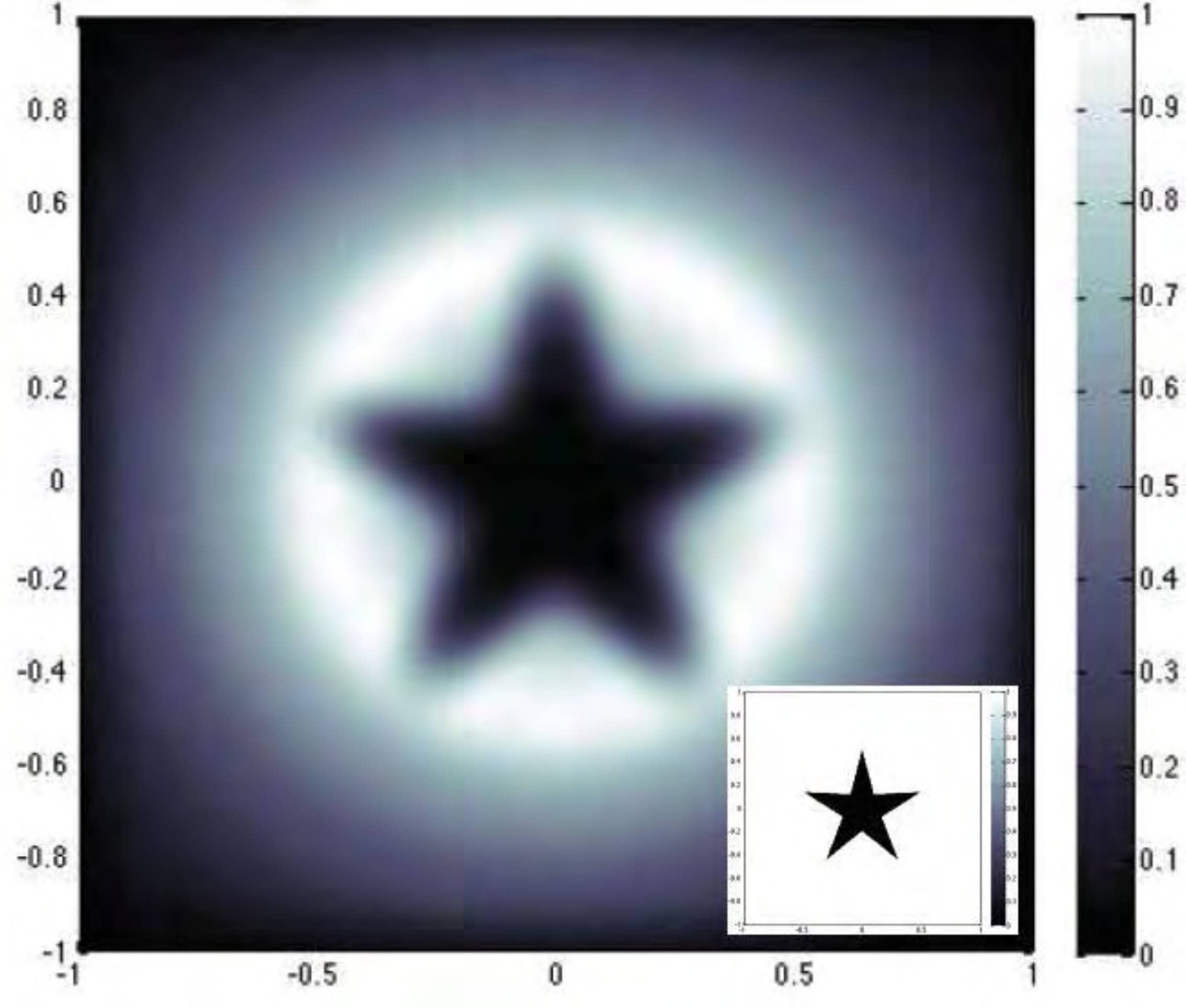}\\(a)
 	\end{minipage}
 	\begin{minipage}[c]{0.48\columnwidth}
 		\includegraphics[height=1.5in]{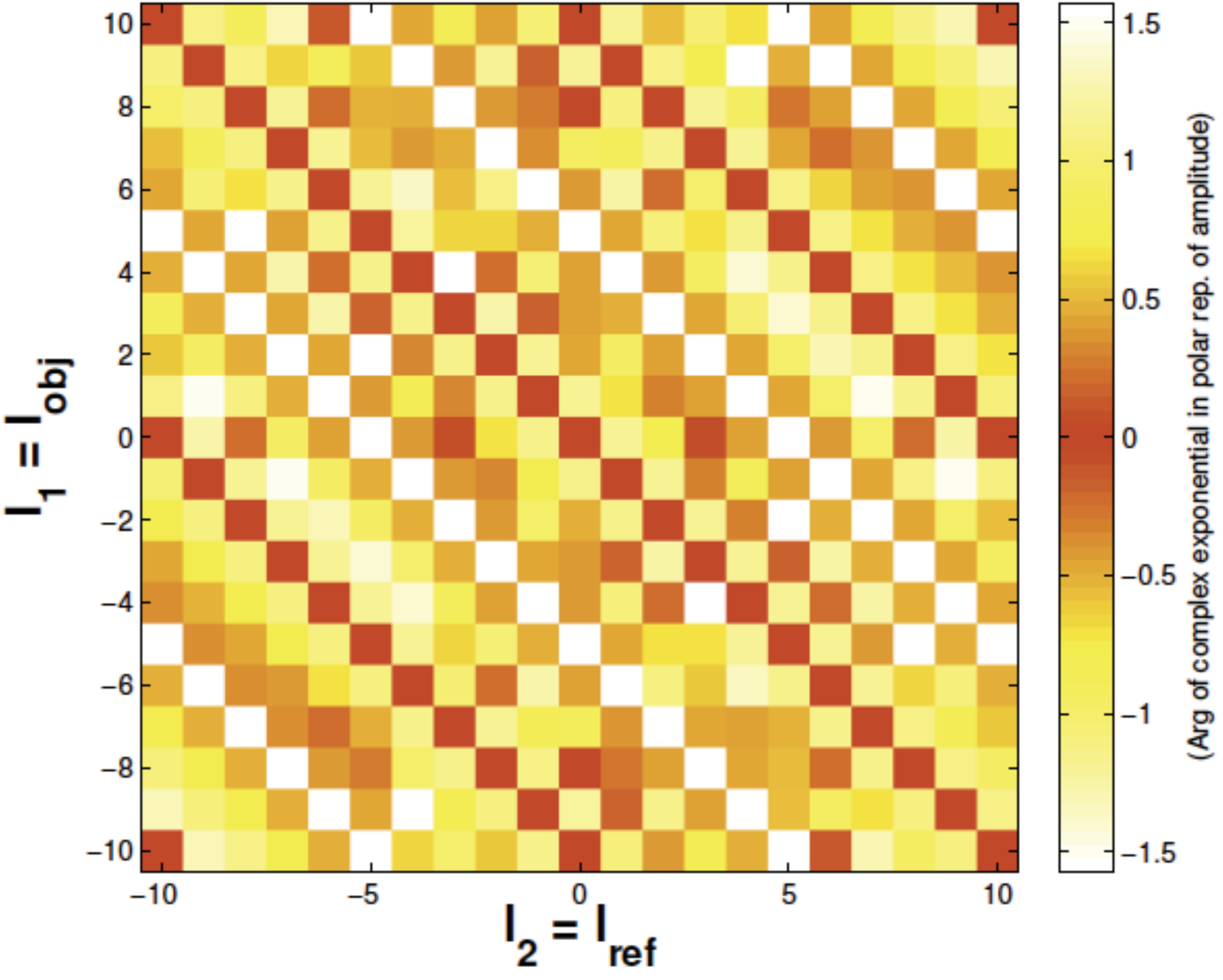}\\(b)
 	\end{minipage}
	\begin{minipage}[b]{0.48\columnwidth}
		\includegraphics[height=1.5in]{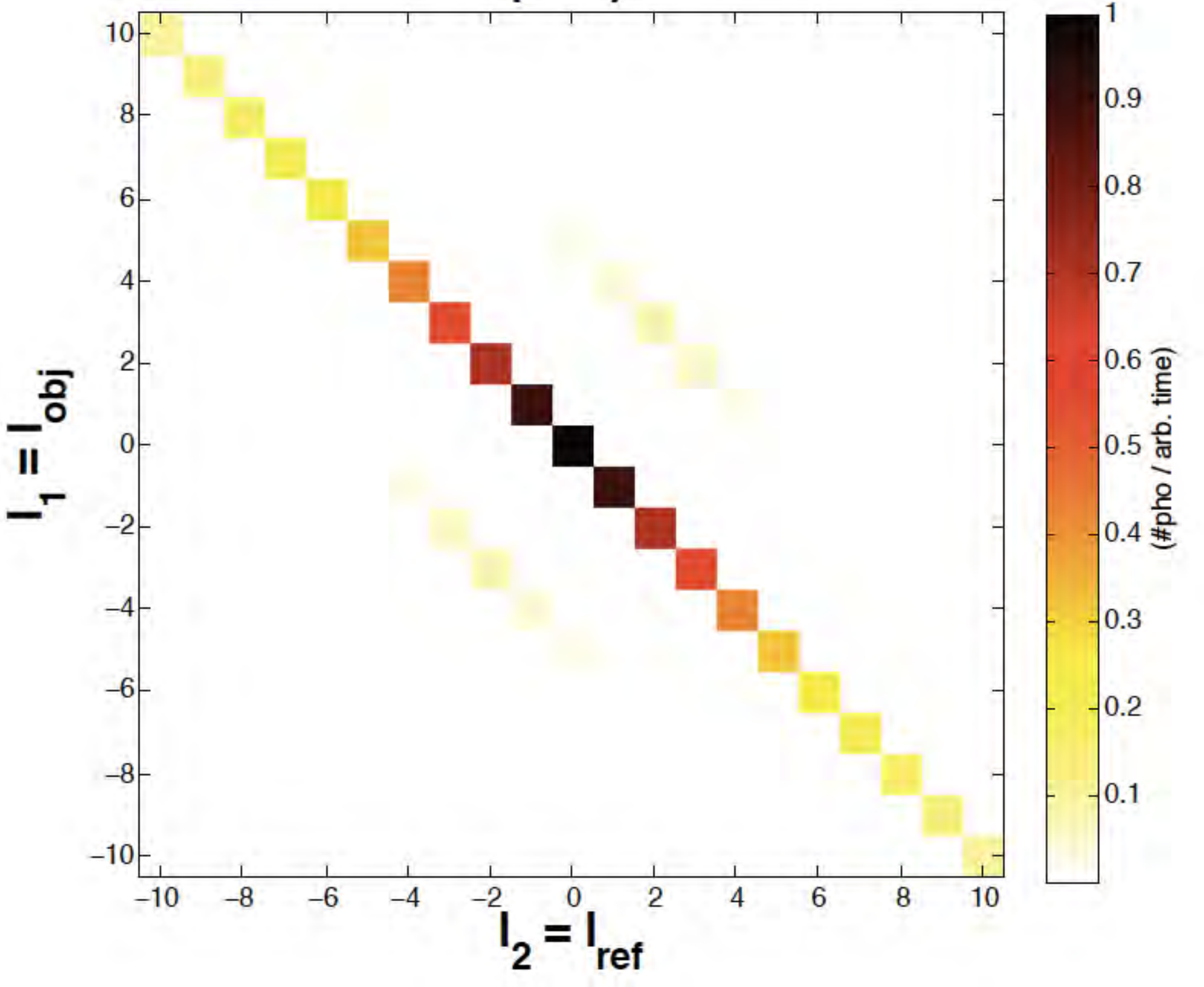}\\(c)
	\end{minipage}
	\begin{minipage}[b]{0.48\columnwidth}
		\includegraphics[height=1.5in]{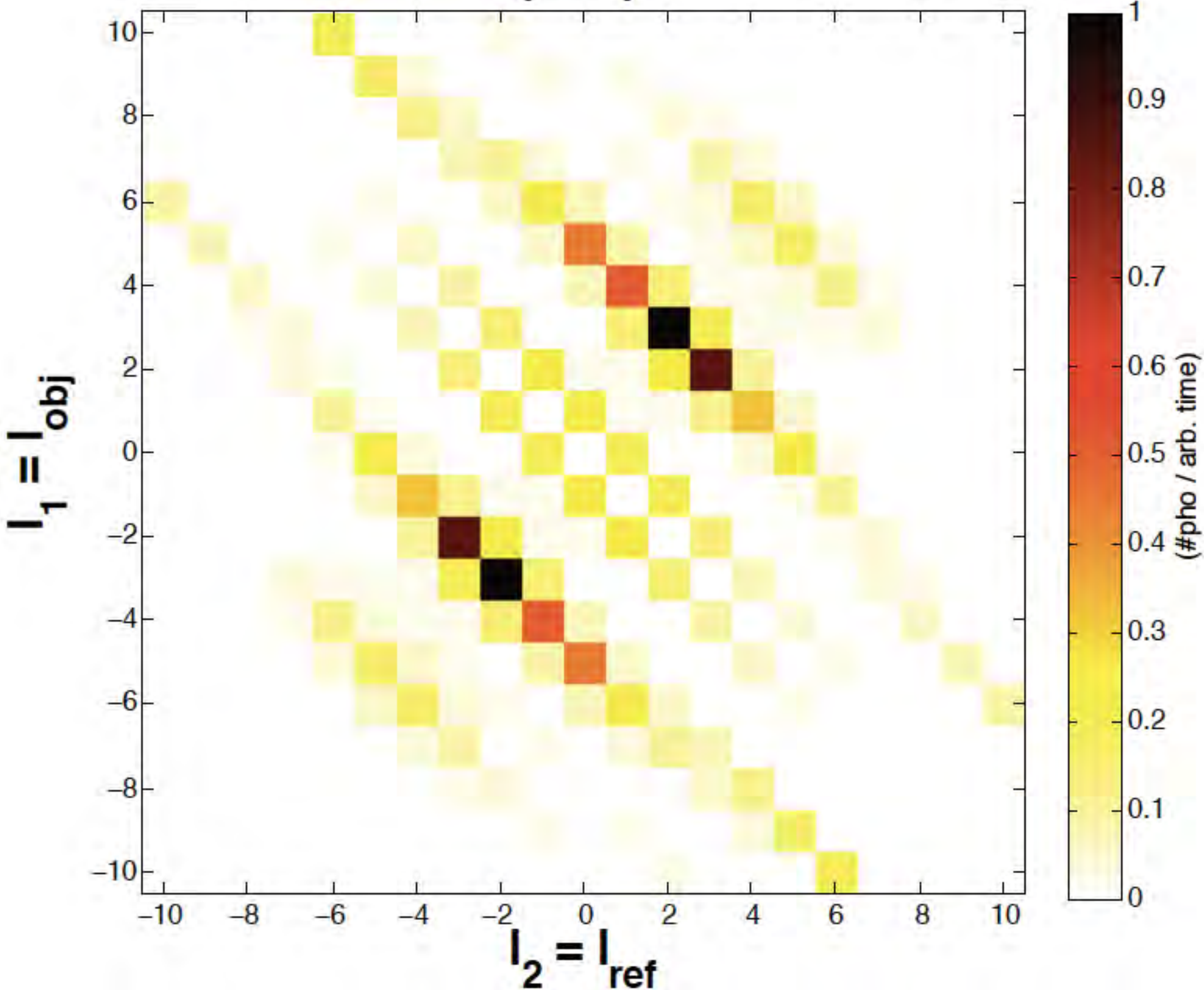}\\(d)
	\end{minipage}
 \caption{(a) The CSI reconstruction using $l_{max}=10$, $p_{max}=7$, with opaque star object of max width $0.9w_0$ in lower right corner. (b) The amplitude phase spectrum of the amplitudes in Eq.~\ref{transamp} across all measured $l$ values (that is, the $\alpha$ in Sec.~\ref{theory}), for the case $(p=2, p'=7)$  (c) The joint OAM spectrum of the star, having summed over all $p$, and (d) the same spectrum with the conservation diagonal removed. }
 \label{starcen}
 \end{figure}

Fig.~\ref{starcen}(c) shows the joint spectrum of a simple 5-pointed opaque star (with $5$-fold rotational symmetry) whose dimensions are confined entirely within the beam. The object's lack of radial extension causes a decrease in magnitude of the $l= \pm 5$ components of the joint spectrum, since the LG modes of higher momentum (higher $l$) do not interact with the object . Consequently, the object's  spectral signature in the off-diagonal components of the joint OAM spectrum becomes visually less obvious. However, as Fig. \ref{starcen}(d) shows, by setting the diagonal components of the joint spectrum to $0$ and rescaling the colormap used to view the spectrum, the off-diagonal contributions become much more visible. Since it is these off-diagonal contributions that carry the extra information upon which the CSI setup is based,  in order to improve the contrast of off-diagonal spectral components, we will zero out the conservation diagonal (states with $l_o=-l_r$) for the remaining object spectra simulated in this report. The image reconstructions will include the contributions of the $l_o=-l_r$ states.

Additionally, note that Fig.~\ref{starcen}(b) shows the coincidence spectra of the phases alone, where the signature $5$-fold symmetry in the diagonals is still present. However, the $l_{o}+l_{r}=N$ are not the only non-zero elements in this case.

\begin{figure}
 	\hspace{0.1in}
 	\begin{minipage}{0.32\textwidth}
		 \includegraphics[width=\linewidth]{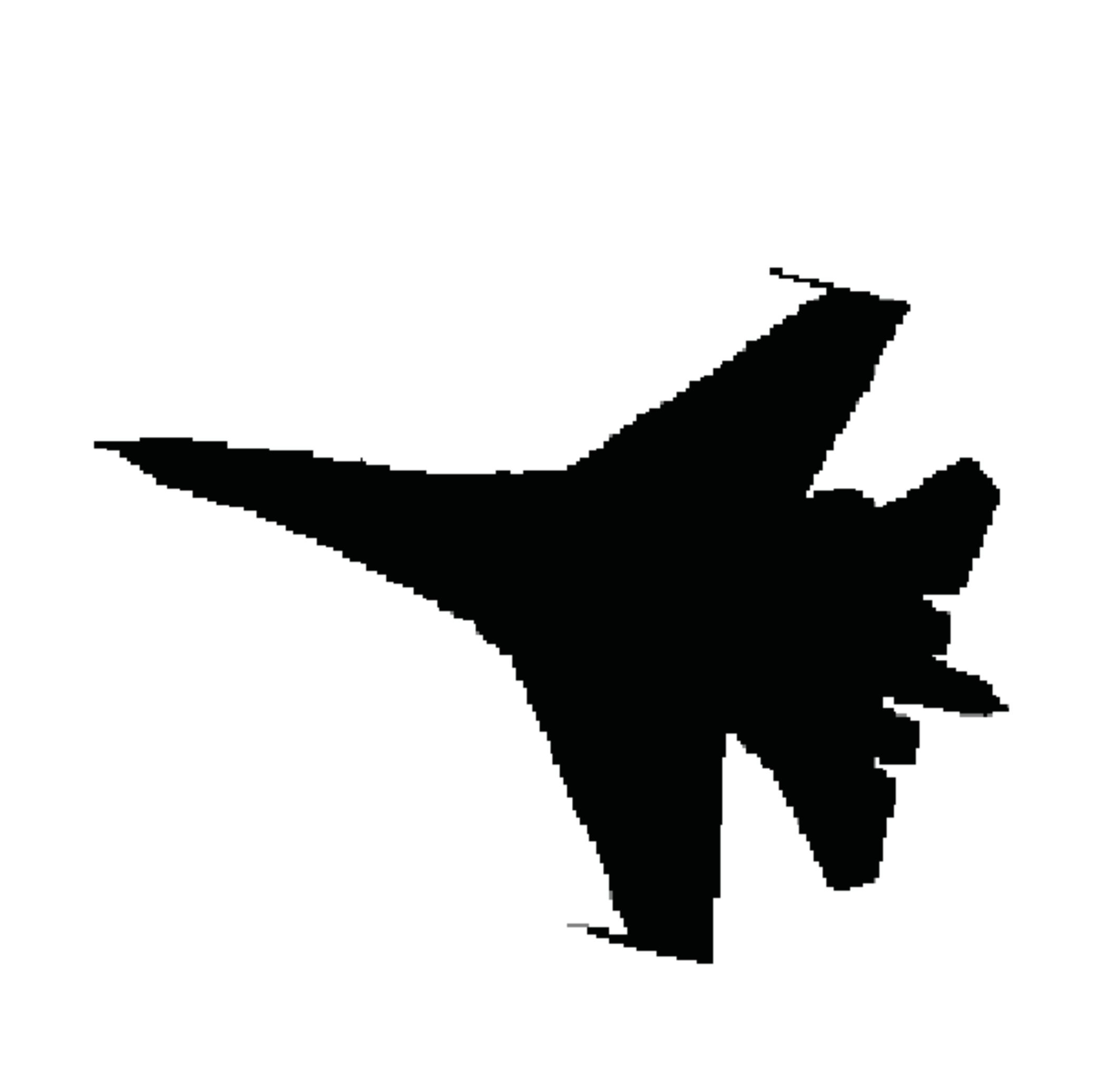}\\(a)
 	\end{minipage}\hfill
 	\begin{minipage}{0.32\textwidth}
 		\includegraphics[width=\linewidth]{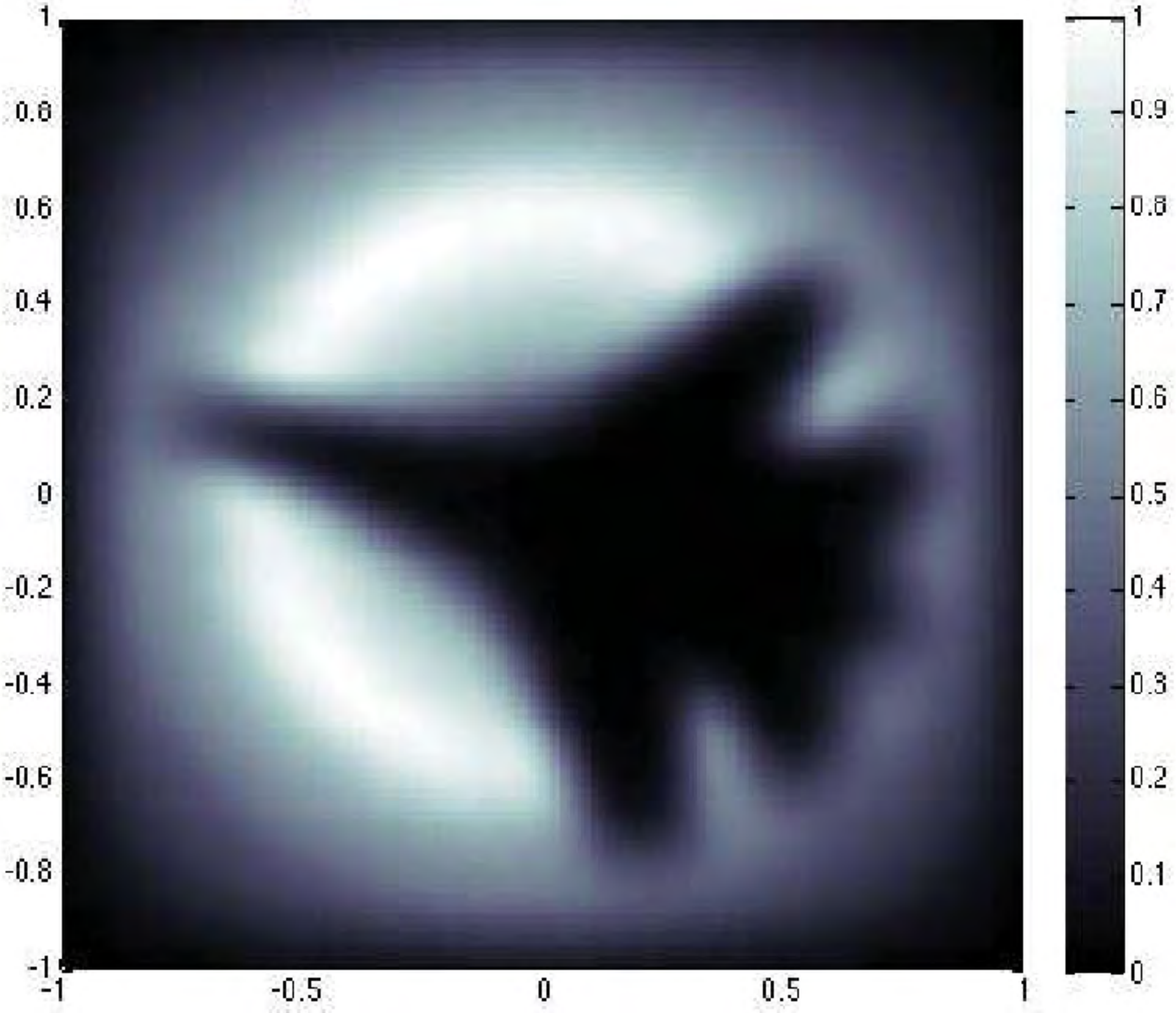}\\(b)
 	\end{minipage}\hfill
	\begin{minipage}{0.32\textwidth}
		\includegraphics[width=\linewidth]{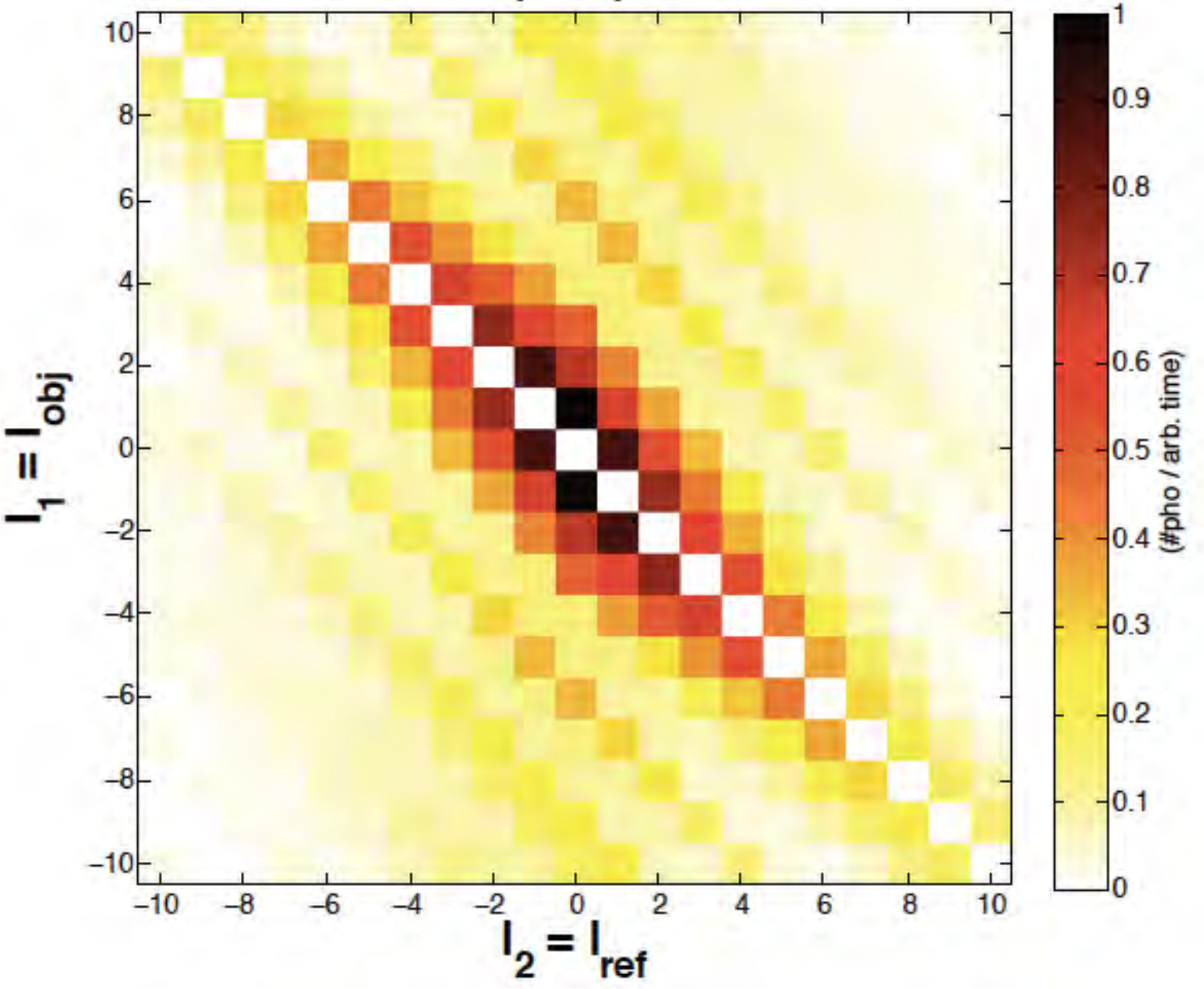}\\(c)
	\end{minipage}\hfill
 \caption{(a) Opaque fighter jet object of max width $w_0$ and, (b) the CSI reconstruction using $l_{max}=10$, $p_{max}=7$; (c) The joint OAM spectrum of the fighter jet, summed over all $p$ with the conservation diagonal removed. }
 \label{fightercen} \end{figure}

\begin{figure}
 	\begin{minipage}{0.32\textwidth}
		 \includegraphics[width=\linewidth]{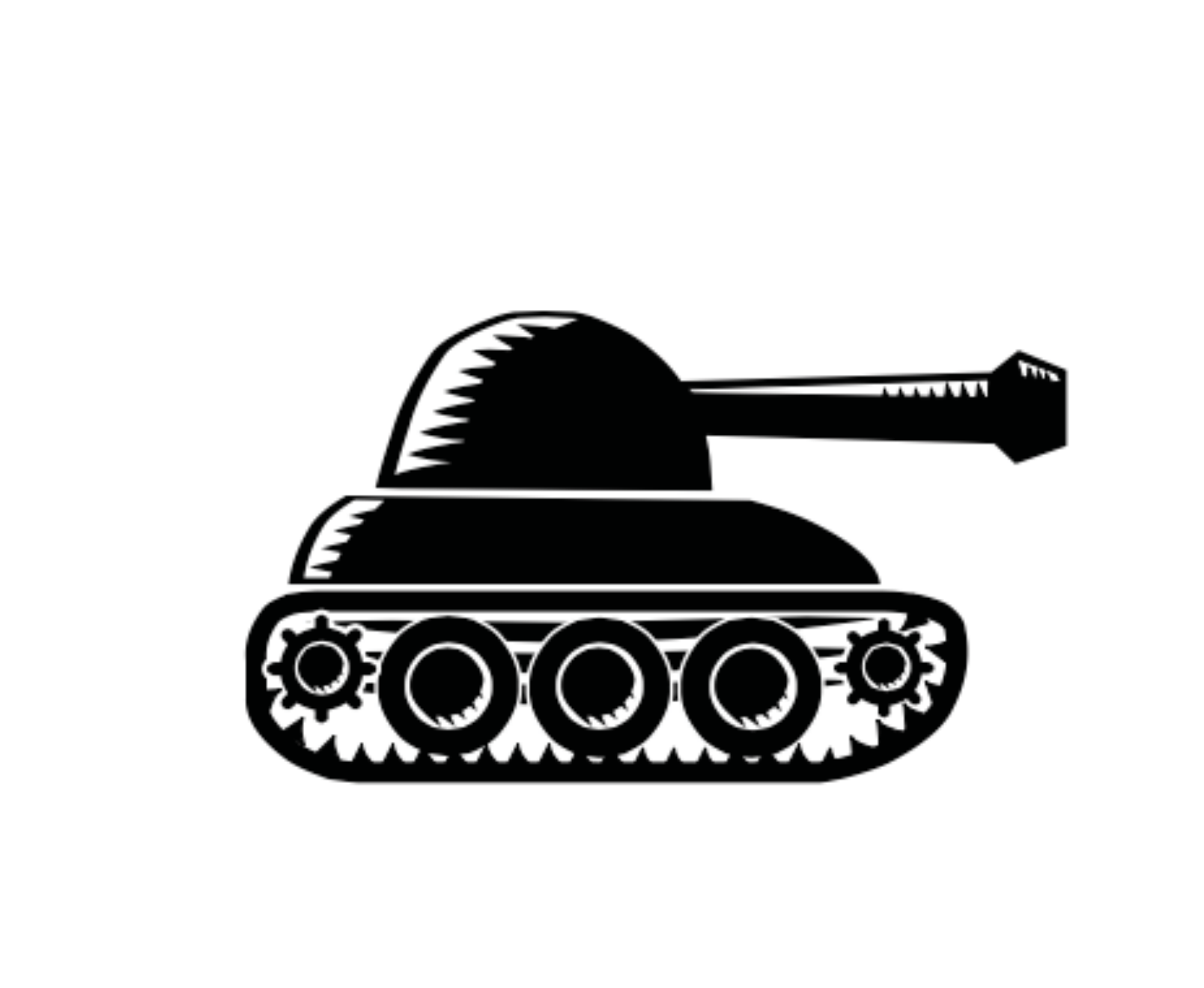}\\(a)
 	\end{minipage}
 	\begin{minipage}{0.32\textwidth}
 		\includegraphics[width=\linewidth]{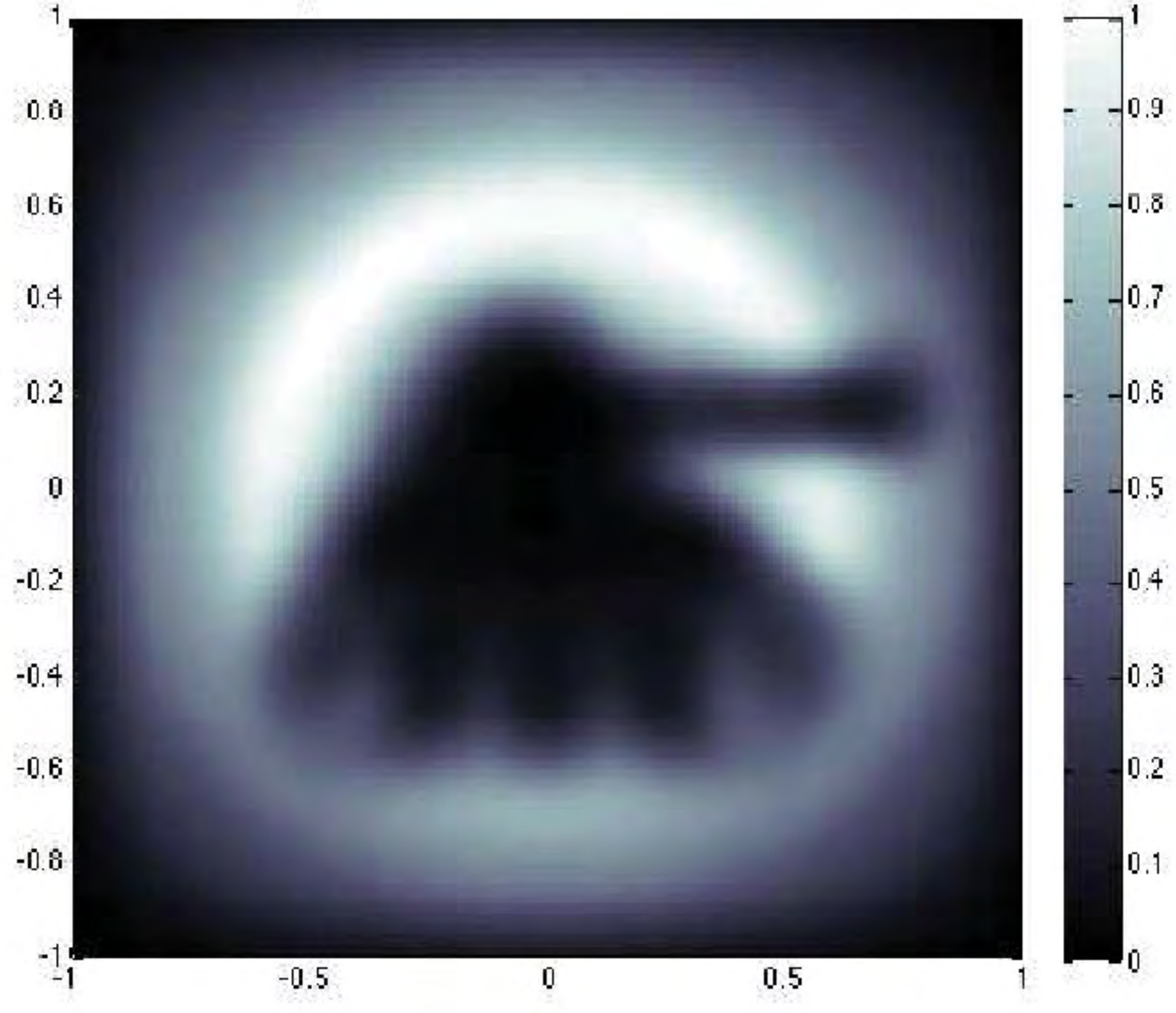}\\(b)
 	\end{minipage}
	\begin{minipage}{0.32\textwidth}
		\includegraphics[width=\linewidth]{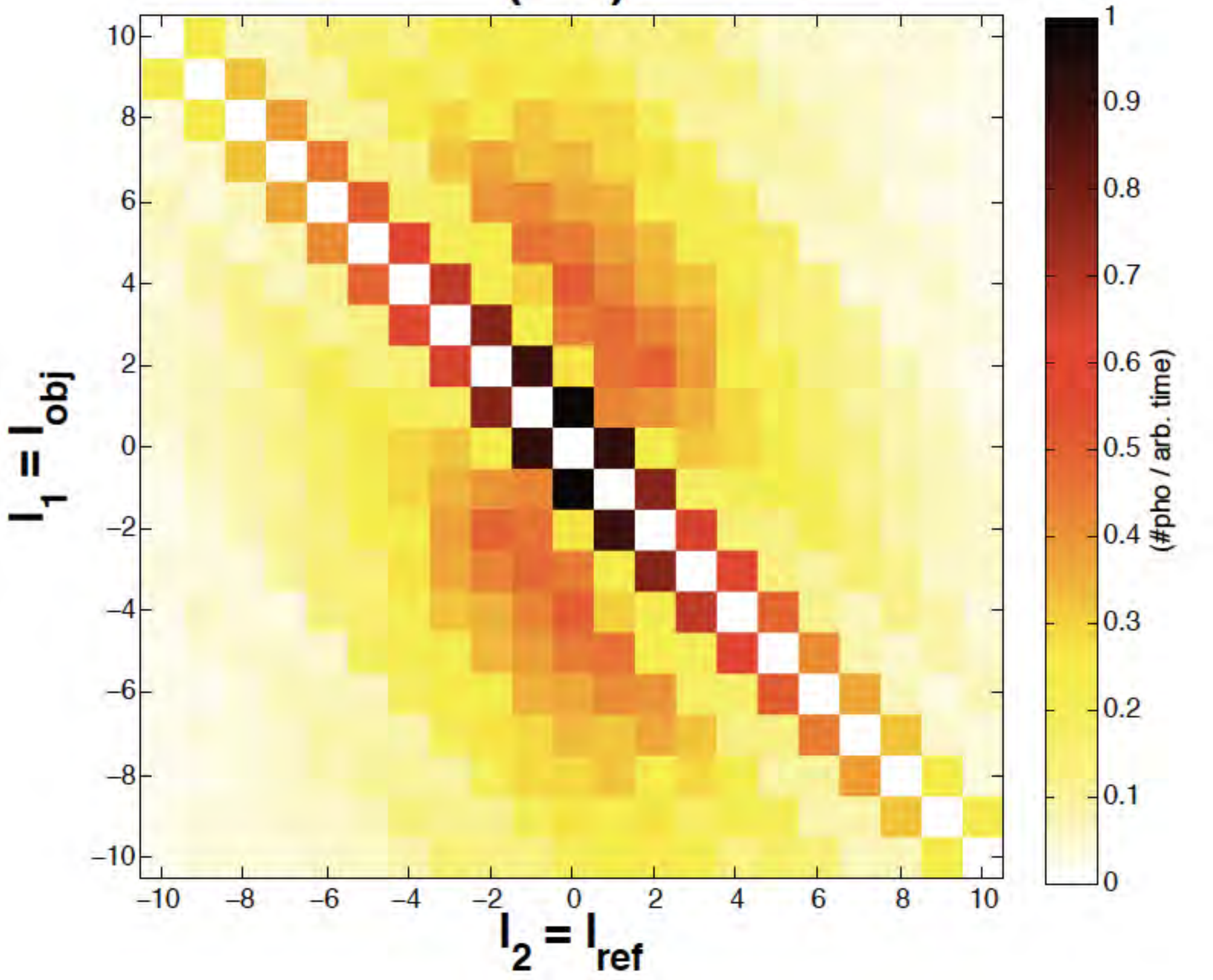}\\(c)
	\end{minipage}
 \caption{(a) Opaque tank object of max width $0.7w_0$ and, (b) the CSI reconstruction using $l_{max}=10$, $p_{max}=7$; (c) The joint OAM spectrum of the tank, summed over all $p$ with the conservation diagonal removed. }
  \label{tankcen}\end{figure}

 Figs. \ref{fightercen}(b) and \ref{tankcen}(b) simulate the ability of the CSI method to image objects with more complicated, less symmetric transmission functions $T(r,\phi)$. The joint spectra shown in Figs. \ref{tankcen}(c) and \ref{fightercen}(c) are clearly less compact than those of simpler objects, like the star. This is to be expected when one views complicated objects as a superposition of many simpler, symmetric objects translated with respect to the beam axis: as shown below (Sec. \ref{offaxis}), translation with respect to the beam axis, even for simple objects, spreads the joint spectrum.

The joint spectra of complex objects (Figs.~\ref{fightercen}, \ref{tankcen}) make clear why it is necessary to look at the {\it entire} joint spectrum, instead of just a slice: For objects that are not perfectly symmetric, the joint spectrum looks different along any given slice. Therefore, to uniquely identify an object, each element of the joint spectrum should be considered.

\subsection{Off-Axis Translation and Mutual Information}\label{offaxis}

\begin{figure}
 	\hspace{0.1in}
 	\begin{minipage}[c]{0.46\columnwidth}
 		\includegraphics[height=1.5in]{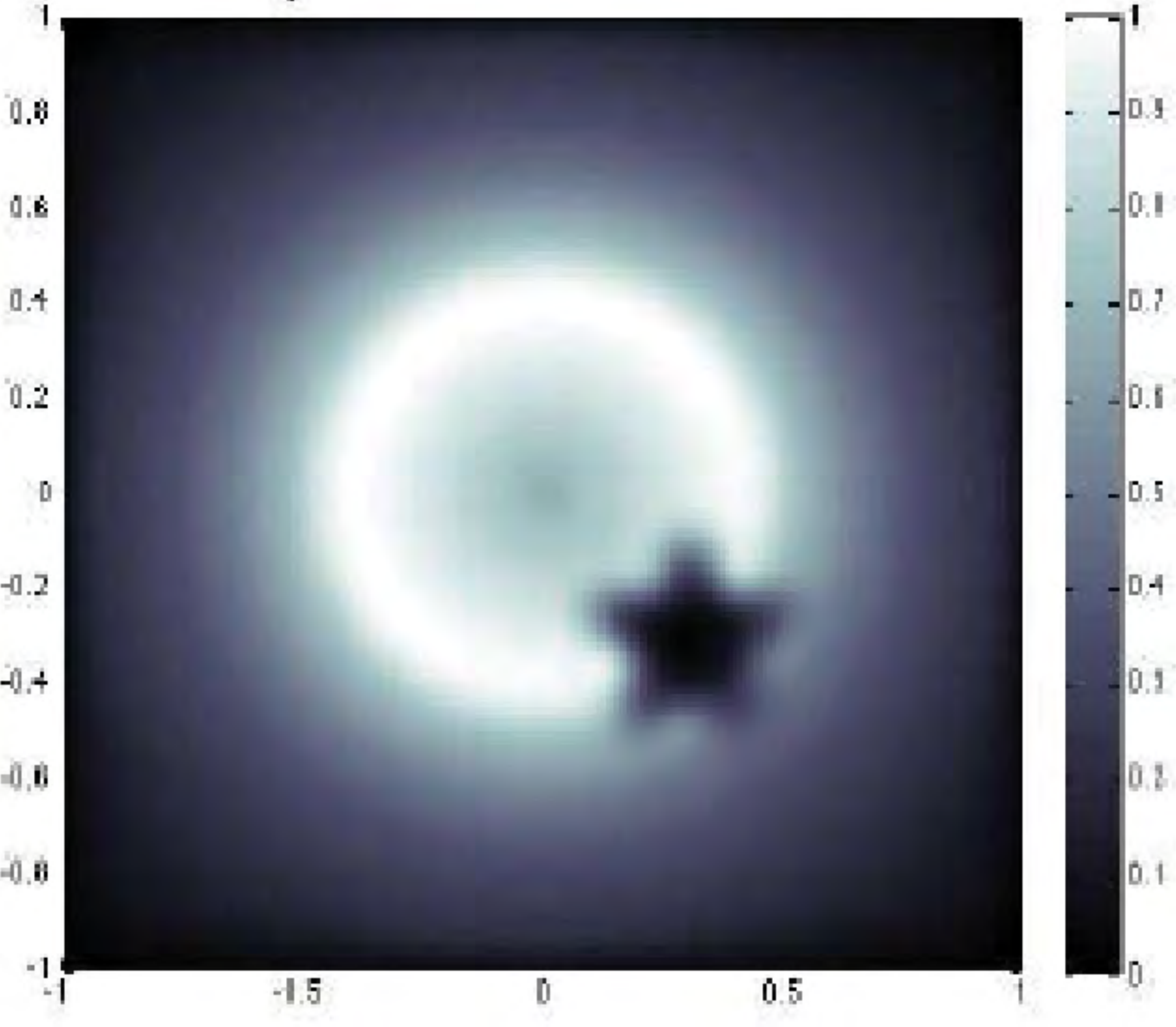}\\(a)
 	\end{minipage}
	\begin{minipage}[c]{0.48\columnwidth}
		\includegraphics[height=1.5in]{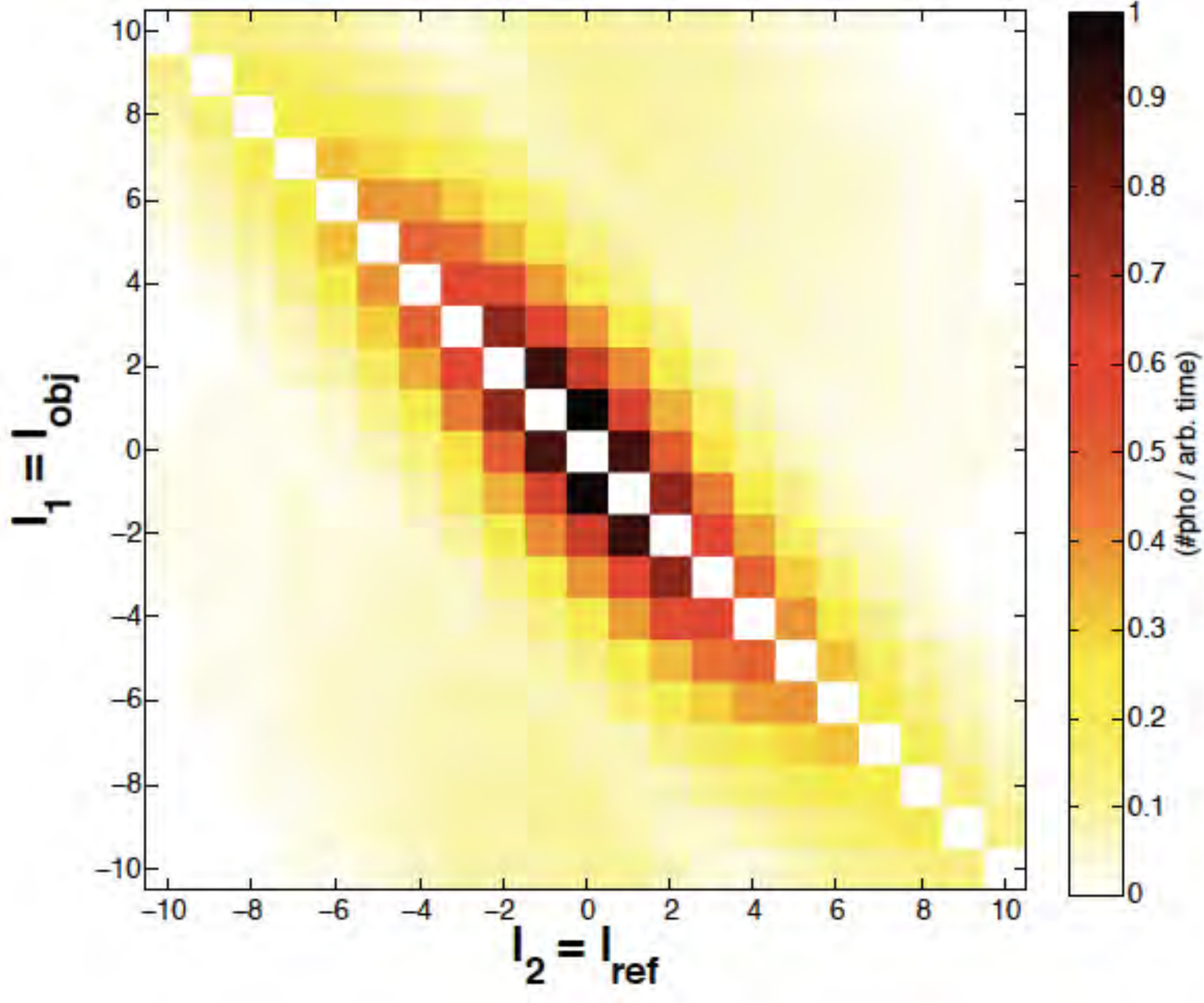}\\(b)
	\end{minipage}
 \caption{(a)  The CSI reconstruction of a translated opaque star, using $l_{max}=10$, $p_{max}=7$; (b) The joint OAM spectrum of the translated star, summed over all $p$ with the conservation diagonal removed. }
 \label{staroff}\end{figure}

\begin{figure}
 	\hspace{0.1in}
 	\begin{minipage}[c]{0.46\columnwidth}
 		\includegraphics[height=1.5in]{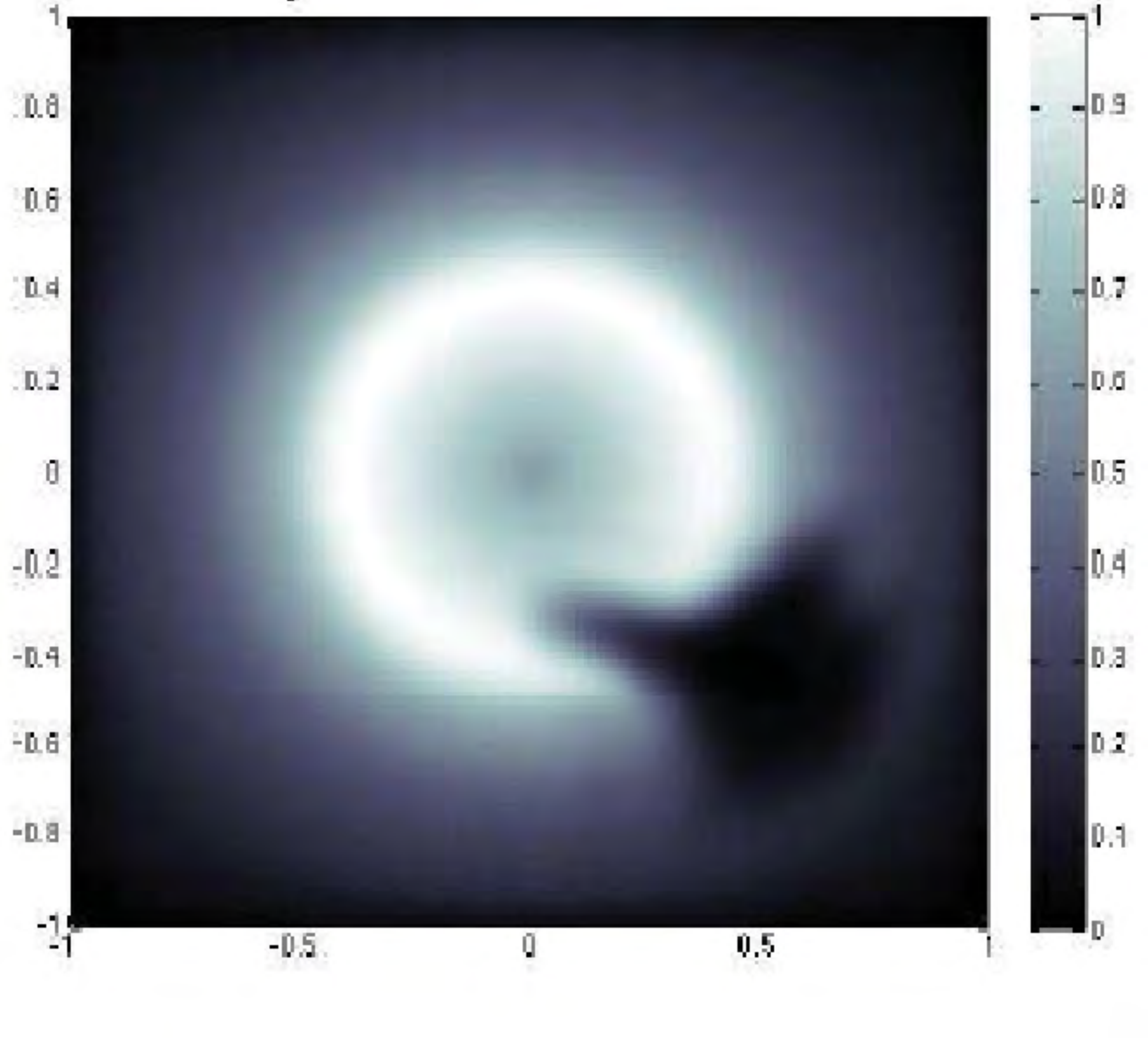}\\(a)
 	\end{minipage}
	\begin{minipage}[c]{0.48\columnwidth}
		\includegraphics[height=1.5in]{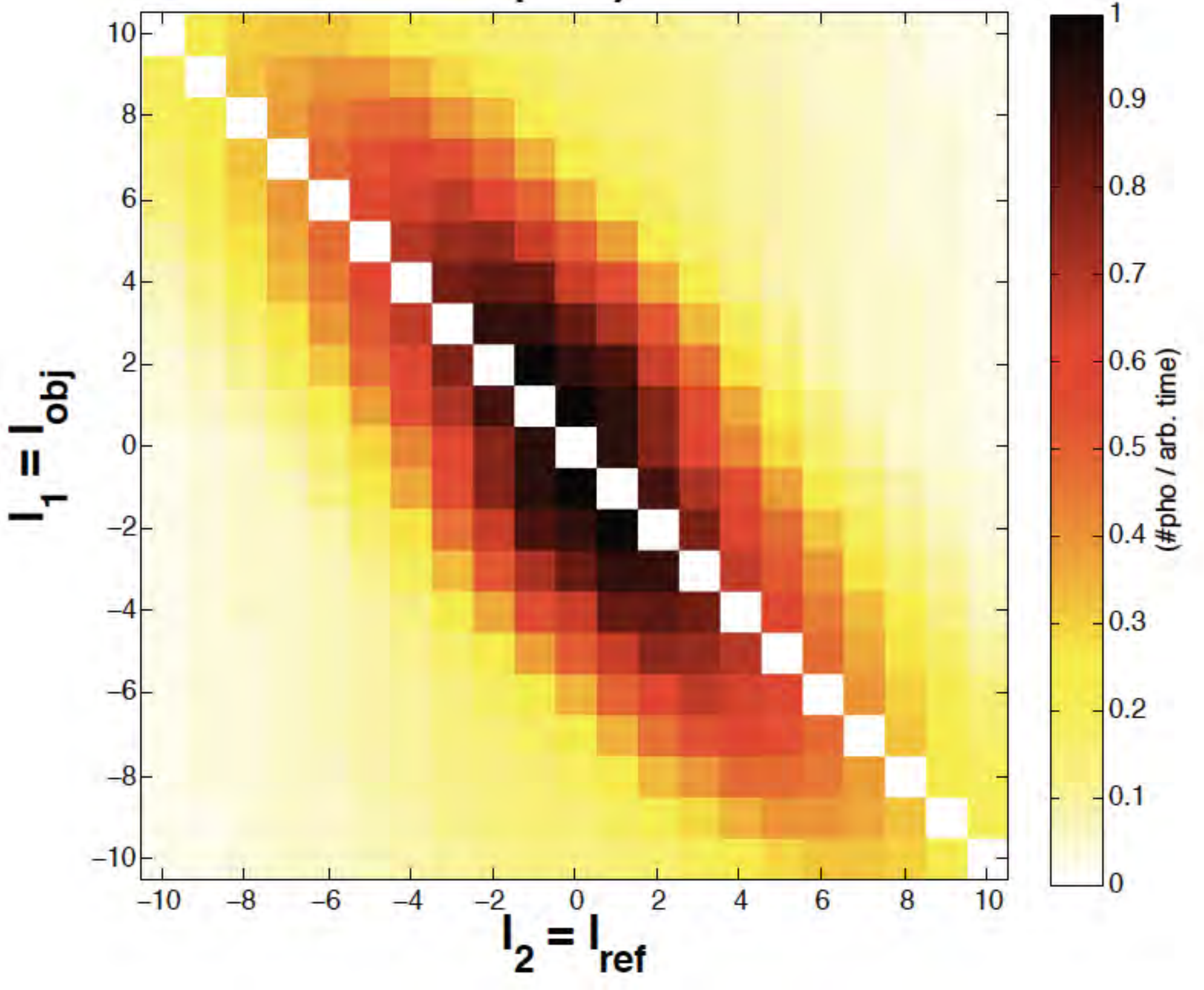}\\(b)
	\end{minipage}
 \caption{(a) The CSI reconstruction of a translated opaque fighter jet, using $l_{max}=10$, $p_{max}=7$; (b) The joint OAM spectrum of the fighter jet, summed over all $p$ with the conservation diagonal removed. }
 \label{fighteroff}\end{figure}

\begin{figure}
 	\hspace{0.1in}
 	\begin{minipage}[c]{0.46\columnwidth}
 		\includegraphics[height=1.5in]{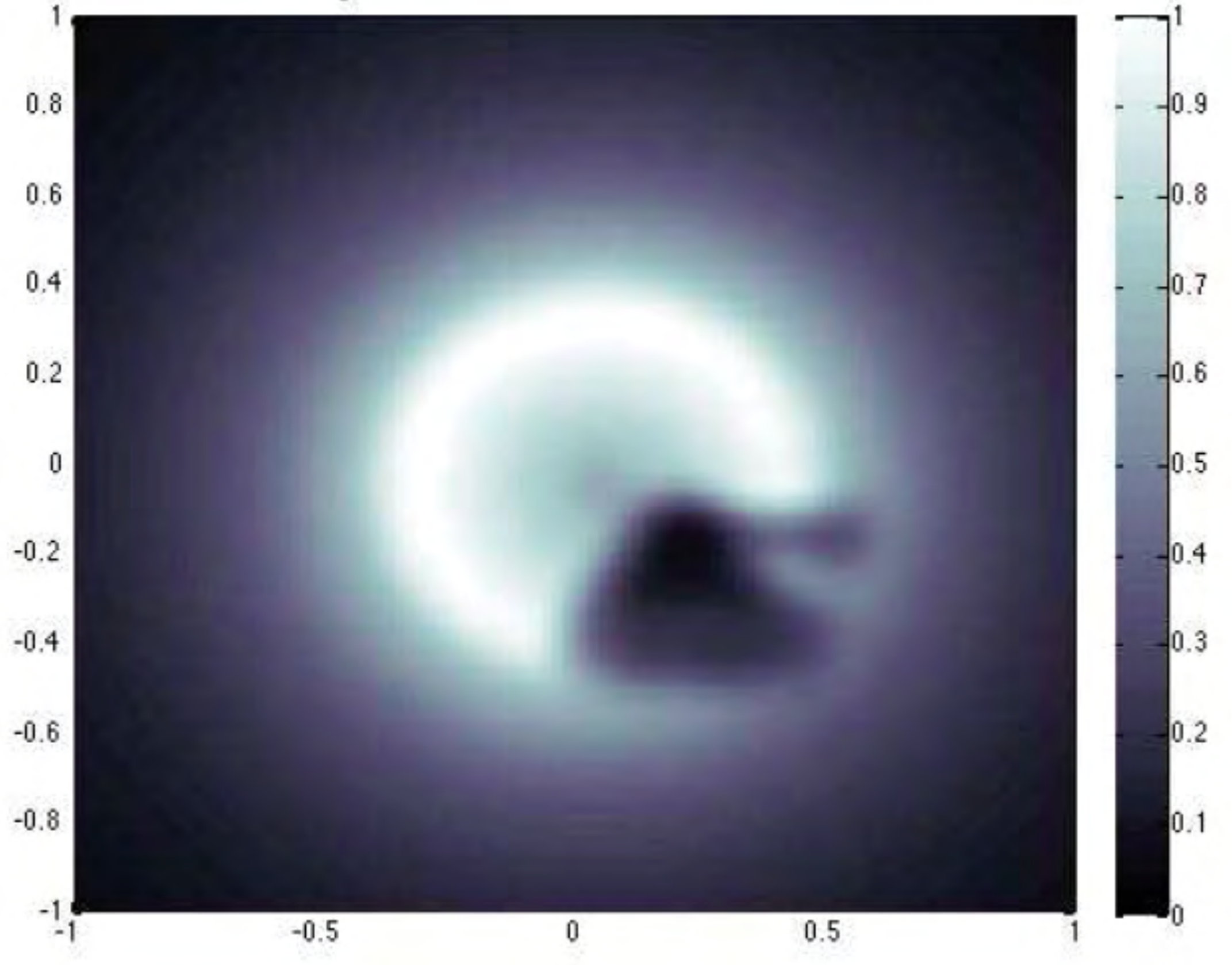}\\(a)
 	\end{minipage}
	\begin{minipage}[c]{0.48\columnwidth}
		\includegraphics[height=1.5in]{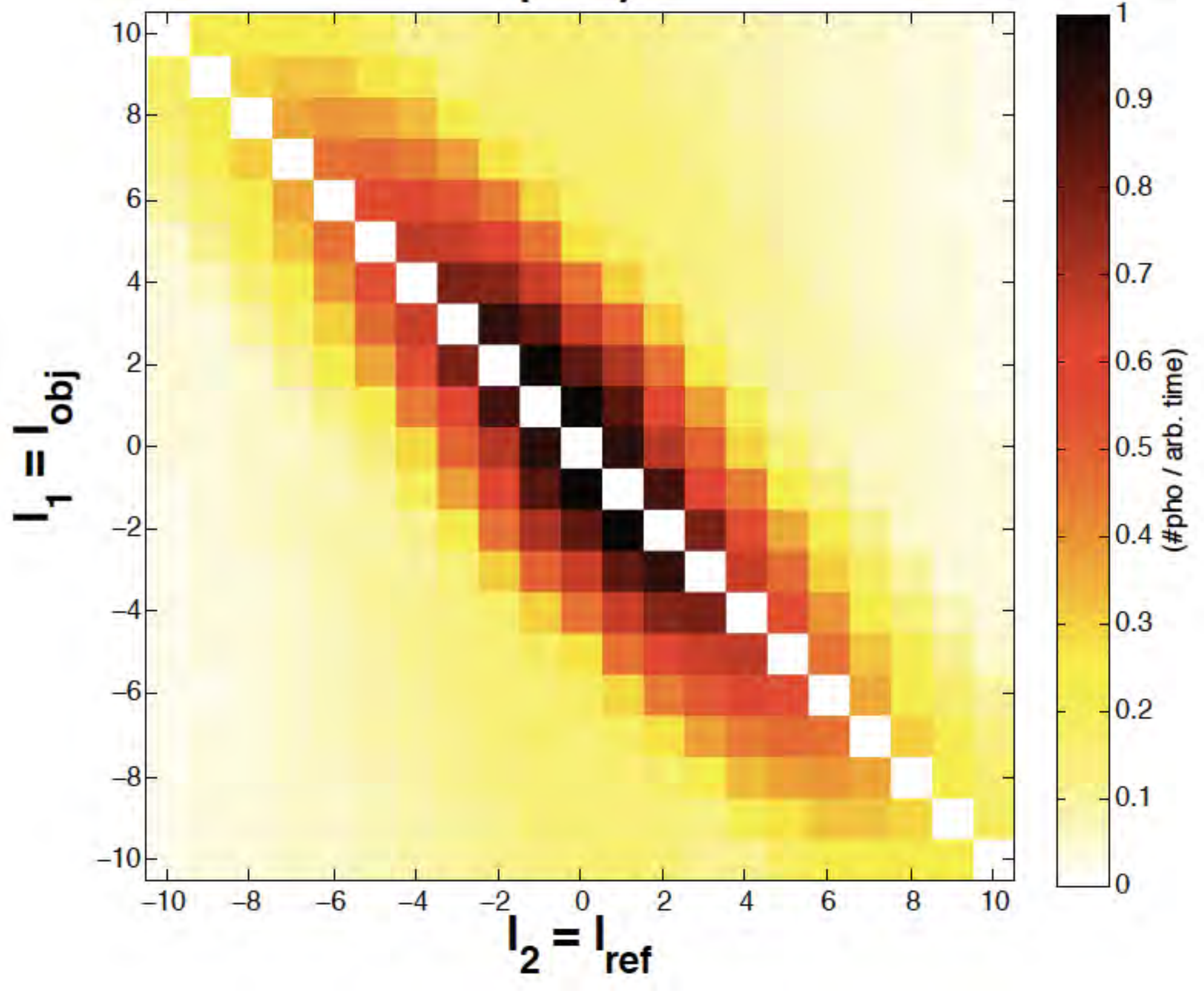}\\(b)
	\end{minipage}
 \caption{(a) The CSI reconstruction of a translated opaque tank, using $l_{max}=10$, $p_{max}=7$; (b) The joint OAM spectrum of the tank, summed over all $p$ with the conservation diagonal removed. }
 \label{tankoff}\end{figure}

For scanning-based applications it is necessary to know how the spectral signatures are altered as the object moves off axis. In Figs. ~\ref{staroff}, ~\ref{fighteroff}, and ~\ref{tankoff} we show the image reconstruction and
spectral signatures of the same objects shown in Figs. ~\ref{starcen}, ~\ref{fightercen}, and ~\ref{tankcen}
respectively {\it after having been shrunk by a factor of $4$ and translated radially with respect
to the beam axis} by approximately $0.7w_0-0.9w_0$. The effect of translation is most obvious in
the case of the star, whose centered spectral signature is quite sparse compared to those of the
tank or fighter jet. Namely, we observe that translation with respect to the beam axis causes a
{\it spreading} in the spectral distribution. Although the exact dynamics of the spectral spread
caused by translation vary from object to object, we note that once the object is sufficiently far
from the beam center -- not surprisingly -- the conservation diagonal is all that remains, all
off-diagonal components going to zero.

\begin{figure}
\centering
\includegraphics[totalheight=2.5in]{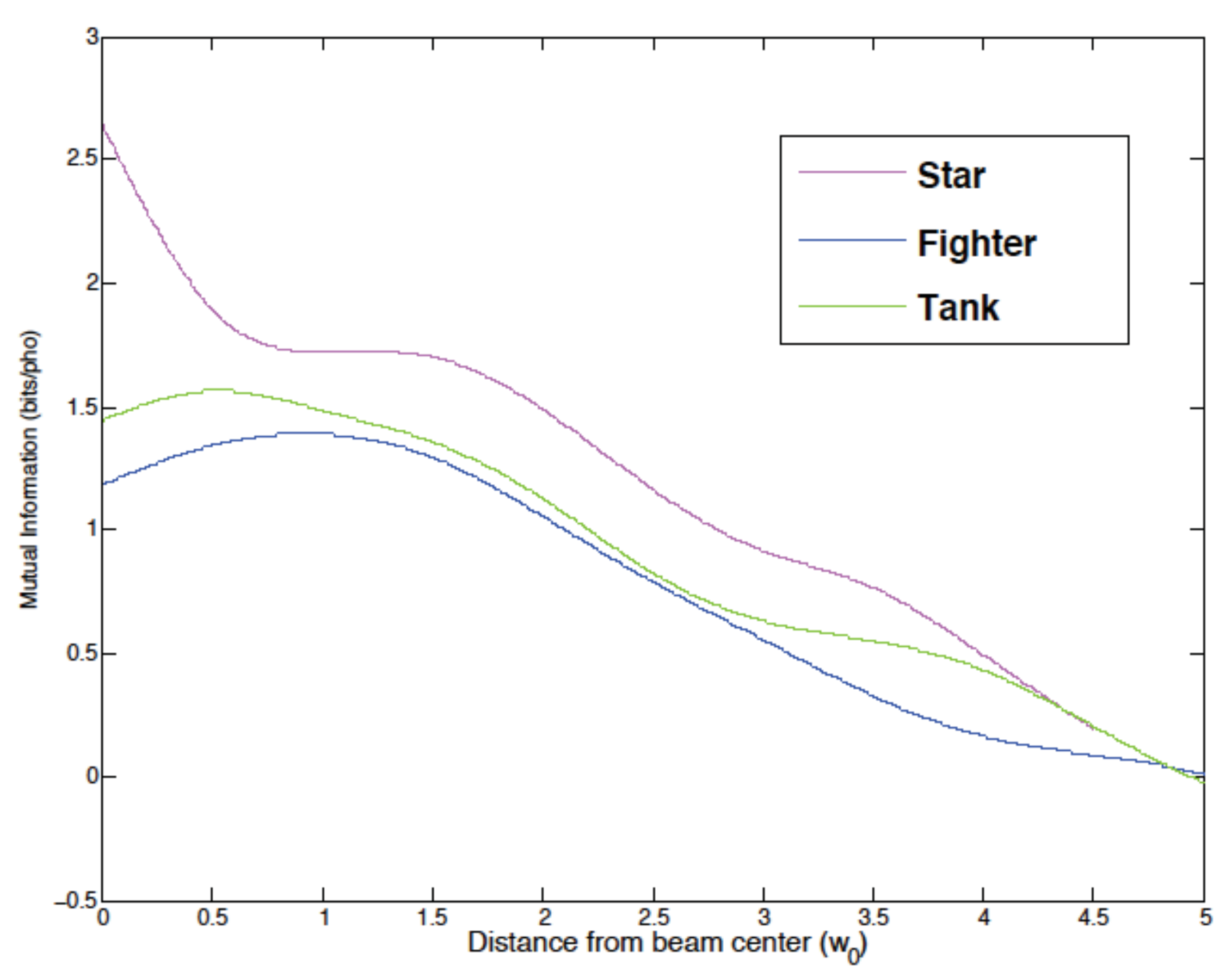}
\caption{Mutual information carried by off-diagonal components of joint OAM spectrum, for various objects, as a function of distance from beam center with $l_{max} = 10, p_{max} = 5$; increasing $p_{max}$ will increase the mutual information substantially. Note that each object's off-diagonal information content exceeds one bit per photon at the beam center.}
\label{mutualinfotrans}\end{figure}

Given the variation in spectral signature as the object is translated through the beam field, we
expect to see a corresponding variation in the mutual information carried by the components of the
joint OAM spectrum. To calculate this change, we simulated the spectra of the above objects several
times, linearly translating them with each iteration, starting from the beam center and ending
effectively outside of the beam field completely. For each position the mutual information was
calculated using Eq.~(\ref{mutinfo}), and the results are plotted as a function of distance in Fig.
~\ref{mutualinfotrans}.  Since we are primarily interested in the information content
of the off-diagonal components of the joint OAM spectrum, we again zero out
the conservation diagonals so that {\it the mutual information calculated represents
information carried exclusively by off-diagonal components of the spectrum}.

We see that {\it even for complex objects near the beam center, the mutual information carried by off-diagonal components of the joint OAM spectrum exceeds one bit per photon}. As expected, the information goes to zero as the object moves sufficiently far from the beam center. Note that the simpler the object here -- the star -- carries the most off-diagonal information, consistent with the argument made in Sec. ~\ref{complexobjects}, that enlarged symmetry groups cause an increase in correlations which in turn causes the mutual information to go up. In fact, as  we increase $p_{max}$ and the objects' symmetries are better approximated, the mutual information for each object goes up. In Fig.~\ref{mutualinfotrans}, $p\in(0,3)$ with the star's $I_{max}\approx 2.6$ bits/pho. Increasing $p_{max}$ to $7$ gives an $I_{max}\approx 3.3$ bits/pho in off-diagonal components.

\subsection{Rotational Insensitivity of Object Signature} 

\begin{figure}
 	\hspace{0.1in}
 	\begin{minipage}{0.32\textwidth}
		 \includegraphics[width=\linewidth]{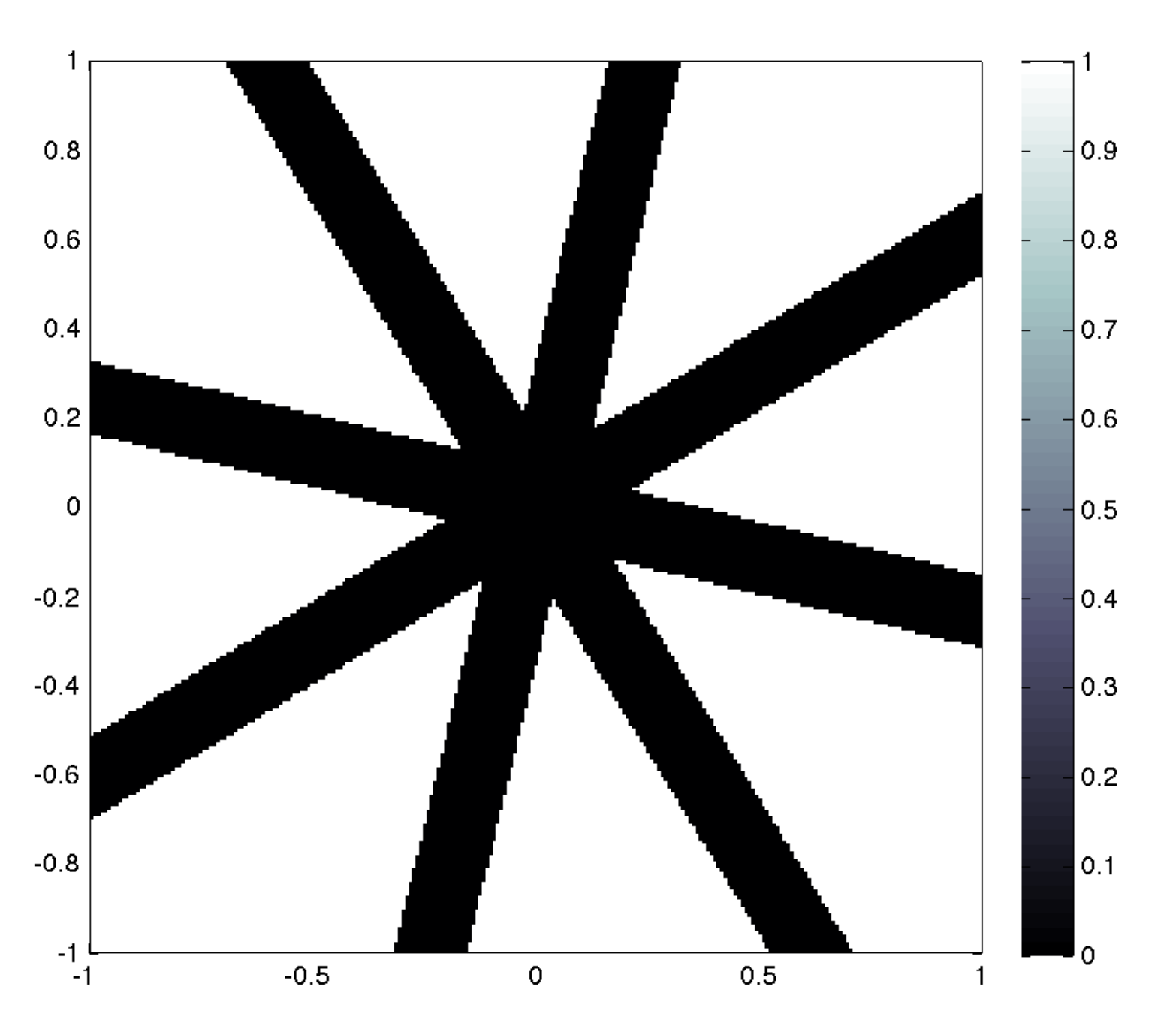}\\(a)
 	\end{minipage}\hfill
 	\begin{minipage}{0.32\textwidth}
 		\includegraphics[width=\linewidth]{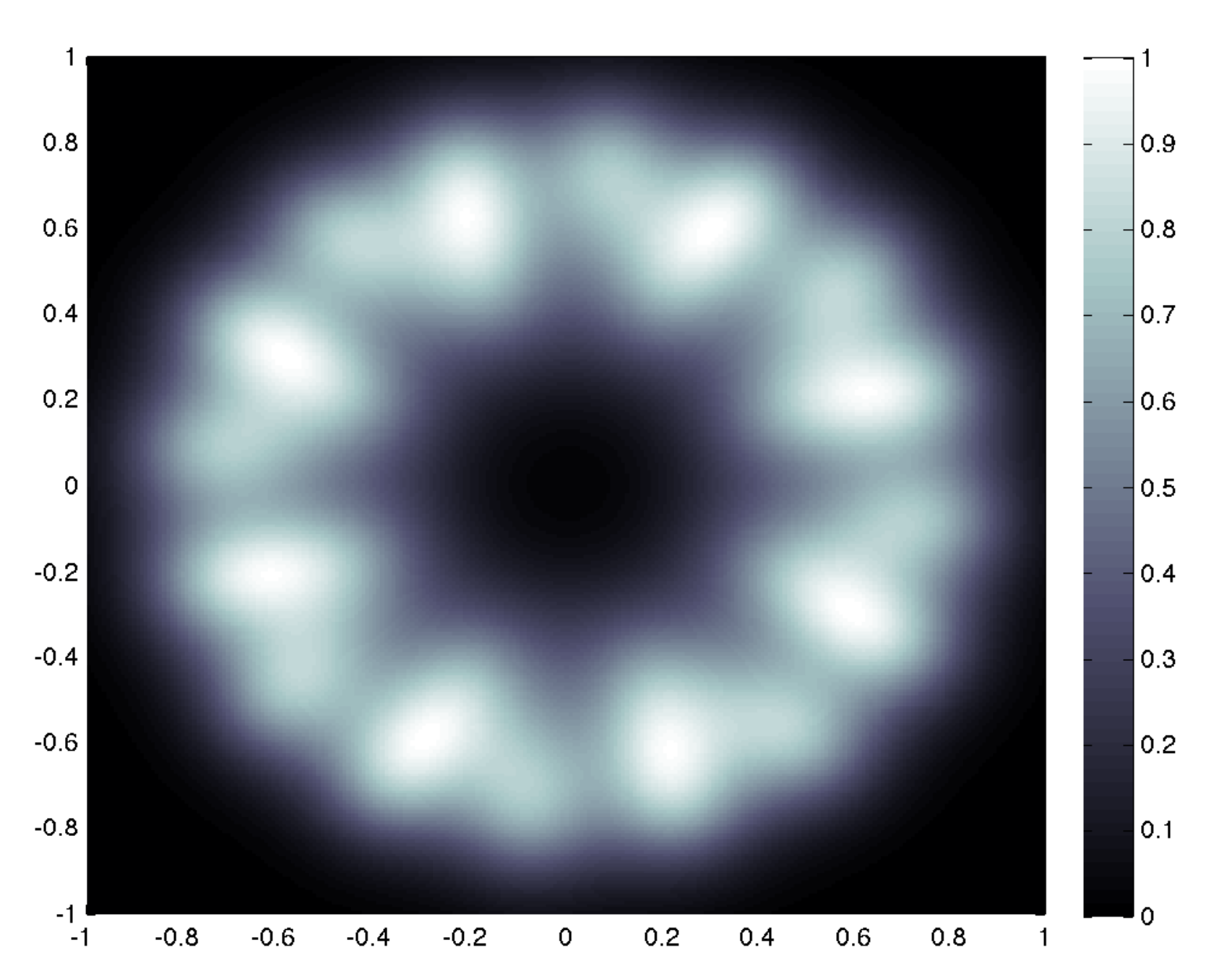}\\(b)
 	\end{minipage}\hfill
	\begin{minipage}{0.32\textwidth}
		\includegraphics[width=\linewidth]{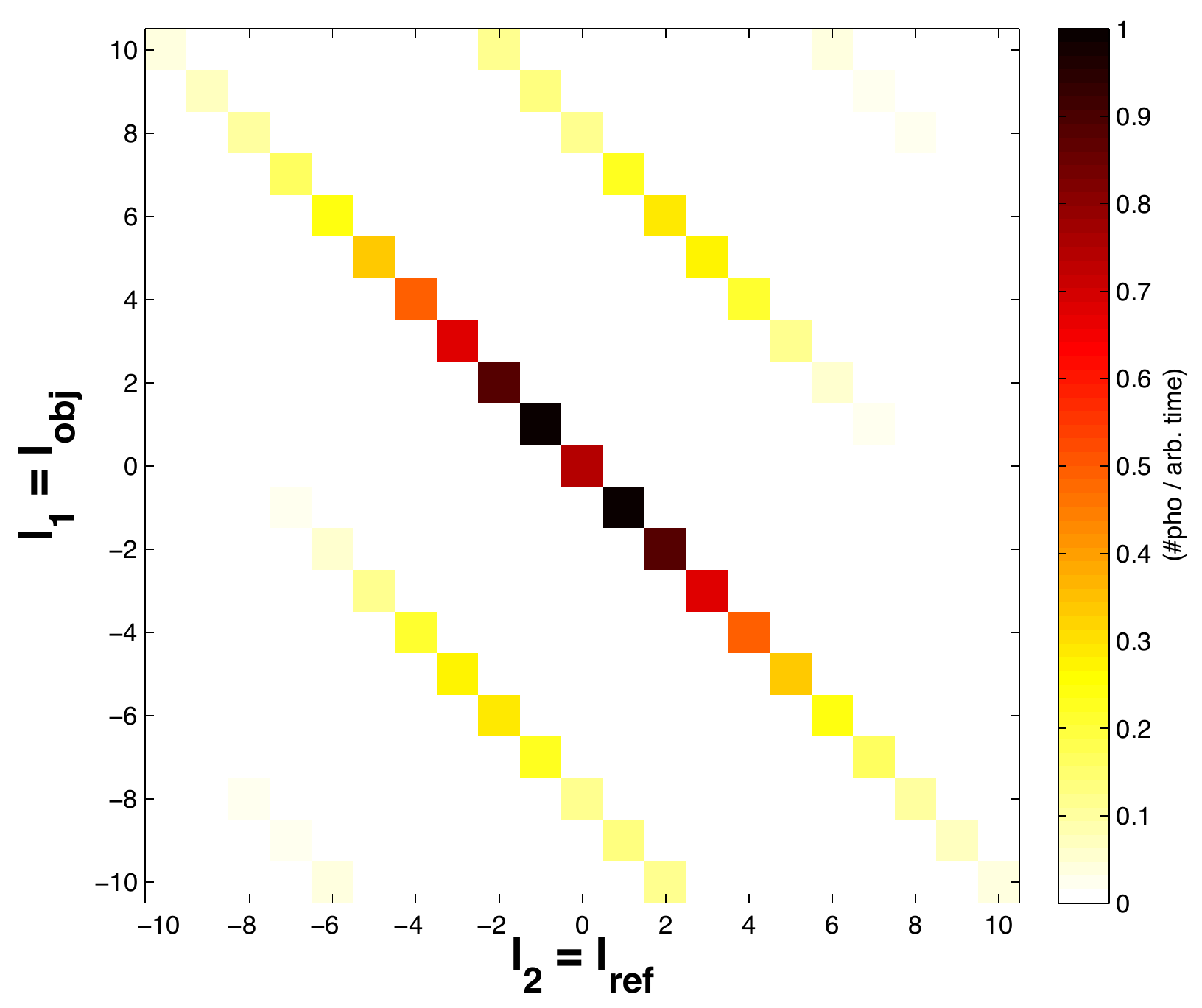}\\(c)
	\end{minipage}\hfill
 \caption{(a)One of many orientations measured for the $8$-fold symmetric object; (b) The CSI reconstruction is blurry as expected, due to the random rotations of the object; (c) The joint OAM spectrum, however, is insensitive to object rotation about a fixed axis. }
 \label{4bandrotation} \end{figure}

Fig.~\ref{4bandrotation} shows an object, image reconstruction, and joint OAM spectrum that was generated following the same procedure described in the rest of this paper, {\it with the addition of randomly rotating the object in between each measurement} (i.e., in between the calculation of each amplitude). Fig.~\ref{4bandrotation}(a) shows the object, an $8$-fold symmetric arrangement of bands, similar to those used in the experiments of \cite{19}. The object is rotated about its central axis by a random angle before each amplitude is calculated. As expected, the image reconstruction profile shown in Fig.~\ref{4bandrotation}(b) is radially blurred about this central axis because the expansion basis vectors that describe the image are being rotated with respect to each other upon each measurement. However, the joint OAM spectrum shown in Fig.~\ref{4bandrotation}(c) is unaffected by the series of random rotations. Because the object looks the same from the OAM mode's frame of reference despite the changing orientation about the central axis, the joint OAM spectrum is as we would expect for an $8$-fold symmetric object. 

This result is particularly promising for future application. For example, imagine the object in \ref{4bandrotation}(a) is a rotating helicopter blade. Standard imaging would likely result in something similar to \ref{4bandrotation}(b). However, the CSI technology allows identification via the spectrum in \ref{4bandrotation}(c), with ease and low energy consumption.

Further, note that {\it only $p=0$ values were considered here}. Thus these results could be achieved in a laboratory or application setting using currently existing technology, since the discrimination of higher-order OAM $p$ values is not needed. When higher-order $p$ values are included, object discrimination should become easier.

\subsection{Discussion} The above simulations demonstrate the informational capacity of off-diagonal
components in the joint OAM spectra. We have  exploited this capacity for the purposes of imaging
and object identification by way of the joint OAM spectral signature. Current experimental
barriers, namely the inability to easily detect $p>0$ modes at the single photon level, present
difficulties in physically implementing the experimental apparatus required to recover the phases
of the amplitudes needed for image reconstruction. However, as our simulations indicate, such an
apparatus would be capable of using the information contained in the off-diagonal components of the
joint OAM spectrum to remote image unknown objects {\it without  any record of the spatial
distribution of the photons measured}. The rotational insensitivity of the method is perhaps the most immediately useful property of the method, and owing to its independence of $p$ values, could be applied with currently existing technology.

\section{Conclusions}\label{concludesection}

All techniques discussed involve \emph{no measurements in position space}. The spectral signatures simulated in the final sections of this letter, especially the spectral property of rotational insensitivity, rely only on coincidence measurements. This means that, especially concerning a set of objects with unique signatures or symmetries, our method can be used to detect the presence or absence of objects in question in relatively few measurements as compared to pixel-by-pixel imaging methods.

A number of novel applications suggest themselves based on the results above.  For example, note
that if the object is rotated, the outgoing OAM states simply pick up an overall phase that does
not affect the joint OAM spectrum. This could be useful, because it allows a rapidly rotating
object to be identified from its OAM spectrum using slow detectors and a small number of measurements. In some circumstances, this may be less expensive and more practical than the use of high speed cameras.

The high mutual information capacity of off-diagonal OAM spectral components also makes our method
well suited for sensing rotational symmetries in few measurements, even for moving and rotating objects. Due to the fragility of OAM
states, the advantages of our setup may best be exploited in small scale biological or production
contexts. For example, the scanning of a biological sample using correlated OAM measurements may
enable efficient detection of the presence or absence of certain structures based on the comparison
of theoretical and observed coincidence rates of off-diagonal spectral components. And, since
objects sufficiently far from the beam center do not affect the coincidence rates, as seen by the
mutual information plots in Fig.~\ref{mutualinfotrans}, we can be confident that a sufficiently small
beam waist will yield accurate spectra. Biological {\it apoptosis} (so-called programmed cell
death) is one context in which the presence or absence of cell symmetries plays an important role,
since apoptotic cells lose their symmetry, and so a change in the distribution of symmetries may indicate a cancerous sample. Sickle cell anemia may provide another avenue for future research, since normal red blood cells have circular symmetry, but sickle cells do not.

The research above furthers the informational analysis of off-diagonal joint OAM spectral components, in addition to demonstrating the full reach of CSI's imaging capabilities. We have seen that not only do these  off-diagonal components carry information that allows image reconstruction, especially in the case of non-symmetric objects, but they do so at rates which can well exceed the bit per photon limit at significant distances from the beam center.

\section*{Acknowledgments}
This research was supported by  the DARPA InPho program through US Army Research Office award W911NF-10-1-0404.

\end{document}